\documentclass[a4paper,11pt]{article}
\usepackage[english]{babel}
\usepackage[dvips]{graphics}
\usepackage{epsfig,psfrag,color,graphicx,here}
\usepackage{xspace,array,delarray,tabularx,calc,pifont,rotating}
\usepackage{wrapfig,units,nicefrac}
\usepackage{amscd,amsmath,amssymb,bbm}%subeqn removed
\usepackage{amsthm} %\begin{theorem}
\usepackage[all]{xy} %Arrow in matrices
\usepackage[pdftex]{hyperref}

\usepackage{subfigure}

\numberwithin{equation}{section} %

\newcommand{\CA}{\mathcal{A}}

\renewcommand{\CD}{\mathcal{D}}
\newcommand{\CE}{\mathcal{E}}
\newcommand{\CF}{\mathcal{F}}

\newcommand{\CV}{\mathcal{V}}

\newcommand{\idm}{\mathbbm{1}} %identitymatrix

\newcommand{\BR}{\mathbb{R}}
\newcommand{\BC}{\mathbb{C}}

\newcommand{\dpar}{\partial}

\newcommand{\e}{{\rm e}}

\newcommand{\by}{{\bar{y}}}

\newcommand{\al}{{\alpha}}

\newcommand{\be}{{\beta}}
\newcommand{\bb}{{\bar\beta}}
\newcommand{\de}{\delta}

\newcommand{\ga}{{\gamma}}
\newcommand{\bg}{{\bar\gamma}}

\newcommand{\Y}{\Upsilon}

\newcommand{\dr}{{\rm d}}

\newcommand{\com}[2]{\left[#1,#2\right]} %Commmutator
\title{}
\author{Thorsten Rahn}
\begin{document}

%S\maketitle

\begin{titlepage}
\setcounter{page}{0}
\begin{flushright}
ITP--UH--16/09\\
\end{flushright}

\vskip 1.8cm
\vspace{2cm}

\begin{center}

{\Large\bf Yang-Mills Equations of Motion for the Higgs Sector of\\[10pt] SU(3)-Equivariant Quiver Gauge Theories}

\vspace{15mm}

{\large Thorsten Rahn}\\[5mm] %${}^{1}$
\noindent {\em Institut f\"ur Theoretische Physik,
Leibniz Universit\"at Hannover \\
Appelstra\ss{}e 2, 30167 Hannover, Germany }
\\[5mm]
{Email: {\tt Thorsten.Rahn@itp.uni-hannover.de}}

\vspace{15mm}

\begin{abstract}
\noindent
We consider $SU(3)$-equivariant dimensional reduction of Yang-Mills theory on spaces of the form $\BR\times SU(3)/H$, with $H$ equals either $SU(2)\times U(1)$ or $U(1)\times U(1)$. For the corresponding quiver gauge theory we derive the equations of motion and construct some specific solutions for the Higgs fields using different gauge groups. Specifically we choose the gauge groups $U(6)$ and $U(8)$ for the space $\BR \times \BC P^2$ as well as the gauge group $U(3)$ for the space $\BR\times SU(3)/U(1)\times U(1)$, and derive Yang-Mills equations for the latter one using a spin connection endowed with a non-vanishing torsion. We find that a specific value for the torsion is necessary in order to obtain non-trivial solutions of Yang-Mills equations. Finally, we take the space $\BR\times\BC P^1\times \BC P^2$ and derive the equations of motion for the Higgs sector for a $U\left(3m+3\right)$ gauge theory.

\end{abstract}

\end{center}
\end{titlepage}

\section{Introduction and summary}
%\section{INTRODUCTION AND SUMMARY}
\addcontentsline{toc}{section}{Introduction and summary}

Yang-Mills equations in more than four dimensions naturally appear in the low-energy limit of superstring theories. Furthermore, natural BPS-type equations for gauge fields in dimensions $d>4$, introduced in \cite{CDFN,BPS2}, also appear in superstring compactifications as the conditions for survival of at least one supersymmetry in low-energy effective field theory in four dimensions \cite{GSW}. Some solutions of Yang-Mills equations on $\BR^d$ were found e.\ g.\ in \cite{fairly-1984,fubini-1985,ivanova-1992,ivanova-1993} but have infinite action for $d>4$ . One possibility for obtaining finite-action solutions for the Yang-Mills equations in higher dimensions is to consider them on spaces of the form $\BR\times G/H$, where $G/H$ is a reductive homogeneous space \cite{preprint1,preprint2,preprint3,yangmillsflow}.
On the other hand, dimensional reduction of the higher dimensional gauge theory appearing in the low-energy limit of superstring theories is necessary and a way of performing a dimensional reduction in our framework is well known. The procedure, referred to as coset space dimensional reduction (CSDR) (see e.\,g.\ \cite{CSDR}), is taking advantage of the fact that homogeneous spaces admit isometries. One can then define a gauge theory on the full space and require the fields to depend on the internal coordinates in such a way that they are invariant under a combined action of $G$-isometries and gauge transformations. Doing this, the Higgs and the gauge sector are unified naturally which is another nice feature of the theory.

In this paper, we investigate the structure of $U(p)$ Yang-Mills theories and the corresponding equations of motion as well as some solutions on spaces of the form $\BR\times G/H$. The factor $\BR$ in the product space stands for one of the four flat dimensions we live in. This is a simplification, which could be generalized to four-dimensional Minkowski space, for instance. The ans\"atze one chooses are $G$-equivariant which implements the dimensional reduction along the coset space. The gauge potential of the theory is given by a connection on a vector bundle associated to a specific principal bundle whose structure group determines the gauge group. If the gauge group $U(p)$ is broken down to $\prod_{i=1}^m U(k_i)$, also the gauge potential on the bundle decomposes in such pieces and in general for each block we get a number of Higgs fields that are responsible for the corresponding breakdown \cite{math.DG/0112160,preprint6}. A physical interpretation of this situation is given in the context of type IIA string theory where we can think of these subbundles to be $k_i$ coincident $D$-branes wrapping $G/H$ and the Higgs fields being open string excitations between neighboring blocks of these $D$-branes \cite{preprint4,rank2quiver,preprint6}. Adding fermions allows to obtain a realistic model in compactification to four dimensions \cite{Szabo:2009-1,Szabo:2009-2}.

What we are looking at, are $U(p)$ gauge theories for different $p$ on the symmetric space $\BC P^2 = \frac{SU(3)}{SU(2)\times U(1)}$ as well as on the non-symmetric space $Q_3=\frac{SU(3)}{U(1)\times U(1)}$. Such theories are equivalent to quiver gauge theories and their $SU(3)$-equivariant ans\"atze for the gauge fields were derived in \cite{preprint5}. Here symmetry breaking takes place and the resulting number of Higgs fields depends on the chosen representation of $SU(3)$ which in our case is determined by the gauge group. First, for $\BC P^2$ we take the ans\"atze for a $U(6)$ and $U(8)$ gauge theory which contains two and four Higgs fields, respectively, and derive the equation of motion for these fields. Second, for $Q_3$ we consider a $U(3)$ gauge theory which involves three Higgs fields and derive the field equations for them using a connection with non-vanishing torsion. In order to obtain solvable equations one needs to choose a specific value for the torsion. Finally, we turn our attention to the product space $\BR \times \BC P^1 \times \BC P^2$, and generalize the equivariant ans\"atze from \cite{preprint1} and \cite{preprint5} to a $U(3m+3)$ gauge theory, where $2m+1$ Higgs fields are involved. Then we derive the field strength and show that the Yang-Mills equations yield a system of $2m+1$ coupled second order differential equations for the Higgs fields. Some novel solutions of these Higgs field equations will also be constructed. It would be interesting to extend the equivariant dimensional reduction technique to ten-dimensional heterotic supergravity with internal six-dimensional coset spaces including a nearly K\"ahler background  (see e.\ g.\ \cite{Louis:2001,Cardoso:2002,Cardoso:2003,Frey:2005,Chatzistavrakidis:2008,Chatzistavrakidis:2009}). It would be also interesting to generalize our solutions to such a general setting.

\section{Quivers and Higgs fields}
%\section{QUIVERS AND HIGGS FIELDS}
\addcontentsline{toc}{section}{Quivers and Higgs fields}

In \cite{yangmillsflow}, theories on spaces of the type $\BR\times G/H$ were considered for the case of $G$ coinciding with the gauge group of the corresponding Yang-Mills theory. In such a case only one scalar field enters into the $G$-equivariant ansatz for a gauge potential. In this paper we are going to consider theories where a breakdown of the original gauge symmetry group takes place and therefore more scalar fields get involved. These fields are interpreted as Higgs fields that are responsible for the corresponding symmetry breaking via the Higgs effect.
Note that $G$-equivariant ans\"atze lead via dimensional reduction to quiver gauge theories
%Note that $G$-equivariant ans\"atze that yield the corresponding dimensionally reduced Yang-Mills theory, are equivalent to quiver gauge theories
\cite{preprint4,rank2quiver,preprint6,preprint5}. Such ans\"atze may become complicated expressions and 
%hence generic ways of writing them down, although they exist \cite{math.DG/0112160,preprint5}, are 
their generic form \cite{math.DG/0112160,preprint5} is not easy to handle. Their explicit form also depends on the chosen representation in the following way. Let $G=SU(3)$ and $\underline{C}^{k,l}$ be some highest weight irreducible representation of $SU(3)$. Then this representation is also a representation for a closed subgroup $H$ of $SU(3)$ which is no longer necessarily irreducible but decomposes as
\begin{equation*}%\label{gen ansatz rep}
 \underline{C}^{k,l}\mid_H = \sum_{i=1}^m \rho_i\,,
\end{equation*}
where $\rho_i$ are irreducible representations of $H=SU(2)\times U(1)$ or $U(1)\times U(1)$.
The number of Higgs fields in our theory then depends on the quiver diagram, containing as many vertices as irreps of $H$ exist, and is determined by the number of maps between these irreps induced by the corresponding lowering operators of $SU(3)$. Therefore a quiver diagram is simply based on the weight diagram of the corresponding $SU(3)$ representation. For the case of $\BR\times \BC P^2$ and $\BR\times Q_3$, with $Q_3:=\frac{SU(3)}{U(1)\times U(1)}$, we consider the case where each arrow stands for exactly one real-valued scalar field and therefore it is clear that the higher quiver representation we choose, the more Higgs fields come into play.

%Generalizations of case (\ref{ansatz1 assumptions})
Let $\CE^{k,l}$ be a rank $p$ Hermitian vector bundle over the space $\BR\times \frac{SU(3)}{H}$, associated to an irreducible representation $\underline{C}^{k,l}$ of $SU(3)$, with the structure group $U(p)$. For the $SU(3)$-equivariant case one can generically write the corresponding associated vector bundle as
\begin{equation}\label{gen ansatz bundle}
 \CE^{k,l}=\bigoplus_{i=1}^m E_i^{\BR}\otimes \CV_i\,,
\end{equation} 
where $E_i^{\BR}$ is a rank $k_i$ bundle over $\BR$ and $\CV_i$ a bundle over $SU(3)/H$ having rank $d_i$ which is also the dimension of the corresponding irrep $\rho_i$ of $H$. The gauge group for such bundles is broken as
\begin{equation*}
 U(p) \longrightarrow \prod_{i=1}^{m} U(k_i)\,,
\end{equation*}
where $\sum_{i=1}^m k_id_i = p$.
%where $k_i$ corresponds to the rank of the bundle $E_i^{\BR}$ over $M_D=\BR$ and $d_i$ to the rank of the bundle $\CV_i$ over $G/H$.
An $SU(3)$-equivariant gauge potential on the bundle (\ref{gen ansatz bundle}) is then given by a block-diagonal part and an off-diagonal one:
\begin{equation}\label{gauge potential splitting}
 \CA = \CA^{\text{diag}} + \CA^{\text{off}}\,.
\end{equation}
The block-diagonal part may be written as
\begin{equation*}
 \CA^{\text{diag}}=\bigoplus_{i=1}^m \CA^i\,,
\end{equation*}
where the size of the blocks $\CA^i$ depends on the dimensions of the $H$-irreps as well as of the rank of the bundle over $\BR$. In our case we consider the bundle $E_i^{\BR}$ over $\BR$ to be of rank $k_i=1$ and the connection $A^i$ on $\BR$ to be flat. Therefore, the part of the connection belonging to this bundle vanishes up to gauge invariance and we find
\begin{eqnarray*} 
 \CA^i &=& \idm_{k_i} \otimes B^i
\end{eqnarray*}
with $B^i$ denoting the connection on the coset part of the product space. Therefore the $\CA^i$ are given by $d_i\times d_i$ matrices $B^i$.

The off-diagonal part can be written as
\begin{equation*}
\CA^{\text{off}~ij} = (1-\de_{ij})~\Phi_{ij}\,,
\end{equation*}
where (no summing over $i,j$) $\CA^{\text{off}~ij}$ is meant to be the $(i,j)$-th block of the gauge connection for $i\neq j$ and zero on the block-diagonal part. The $\Phi_{ij}$ on the right hand side corresponds to the specific map that connects the $i$th and the $j$th vertex of the quiver and is the tensor product of maps between the corresponding bundles. This means
\begin{equation*}
 \Phi_{ij} = \phi_{ij}\otimes\beta_{ij}\,,
\end{equation*}
where $\be_{ij}$ are maps connecting two $H$-irreps $\rho_i$ and $\rho_j$ containing the left-invariant basis of one-forms on the coset space, and $\phi_{ij}$ are size $k_i\times k_j$ Higgs fields depending only on the coordinate $\tau$ on $\BR$ in $\BR\times SU(3)/H$. For our consideration, as mentioned above, these are just real-valued scalar fields, one for each arrow of the quiver.

To sum up, we are dealing with an $SU(3)$-equivariant associated vector bundle $ \CE^{k,l}$ over $\BR\times G/H$ defined as
\begin{equation}\label{general associated principal quiver bundle}
 \CE^{k,l}=\bigoplus_{i=1}^m P(\BR\times G/H,U(d_i))\times_{\rho_i} V^i \to \BR\times G/H\,,
\end{equation}
where $G=SU(3)$ and $V^i$ are finite dimensional representation spaces for the representations $\rho_i$ of the subgroup $H$ of $SU(3)$. Each term comes along with the structure group $U(d_i)$ and hence the overall structure group is given by 
\begin{equation}\label{general structure group}
U\left(\sum_{i=1}^m d_i\right),
\end{equation}
obviously depending on the chosen representation of $SU(3)$.

The explicit construction of the quivers, their representation and the underlying $SU(3)$-equivariant gauge theories was done in \cite{preprint5}. It includes also the explicit formulae of the gauge potential and the field strength for spaces of the form
\begin{equation*}
 M_D\times G/H\,,
\end{equation*}
where $G/H$ is either $\mathbb C P^2=\frac{SU(3)}{S(U(2)\times U(1))}$ or $Q_3=\frac{SU(3)}{U(1)\times U(1)}$ and $M_D$ is some manifold of real dimension $D$ endowed with a Riemannian or Lorentzian metric. Futrthermore the $SU(3)$ action has to be trivial on $M_D$. We are going to use these results for our specific cases in order to derive the equations for the Higgs fields of these quiver gauge theories and to give some explicit solutions.
\section[Yang-Mills theory on R x CP2 in C(2,0) quiver representation]{Yang-Mills theory on $\mathbb R \times \BC P^2$ in $\underline{C}^{2,0} $ quiver representation}
%\section{YANG-MILLS THEORY ON $\mathbb R \times \BC P^2$ IN $\underline{C}^{2,0} $ QUIVER REPRESENTATION}
\addcontentsline{toc}{section}{Yang-Mills on R x CP2 in C(2,0) quiver representation}
\label{yang mills on CP^2 in C20 quiver representation}

\paragraph{Invariant 1-forms on \boldmath{$\mathbb C P^2$}.}
\label{Invariant 1-forms on CP2}
First, we want to summarize all the ingredients from \cite{preprint5} that we will need for writing down an $SU(3)$-equivariant ansatz for the gauge potential. It is quite convenient to do all the calculations in the invariant basis of the corresponding space because we can choose our metric to have constant coefficients in this basis. Hence the covariant derivative with respect to this metric is only depending on the structure constants of $SU(3)$. The projective plane $\mathbb C P^2$ is a complex manifold and therefore we can choose local complex coordinates $(y_1,y_2,\bar y_1,\bar y_2)$.
Using them one can write down the invariant 1-forms as \cite{preprint5}
\begin{subequations}
\begin{eqnarray}
\bar\beta&:=&\begin{pmatrix}\bar\beta^{\bar1}\\[4pt]\bar\beta^{\bar2}\end{pmatrix} \qquad
\mbox{with} \qquad \bar\beta^{\bar e}=\frac1{\gamma}~\dr\bar y^{\bar e}-\frac{\bar y^{\bar e}}
{\gamma^2\,(\gamma+1)}\,y^d~\dr\bar y^{\bar d} \ , \label{left invariant cp2 1}\\[4pt]
\beta&=&\begin{pmatrix}\beta^1\\[4pt]\beta^2\end{pmatrix} \qquad
\mbox{with} \qquad \beta^e=\frac1{\gamma}~\dr y^e-\frac{y^e}
{\gamma^2\,(\gamma+1)}\,\bar y^{\bar d}~\dr y^d \ . \label{left invariant cp2 2}
\end{eqnarray}
\end{subequations}
where
\begin{equation}
\gamma~:=~\sqrt{1+Y^\dag\,Y}\,, \qquad
Y=\left(\begin{matrix} y_1\\y_2 \end{matrix}\right)\,. \label{gammaYdef}
\end{equation}

We denote the coset subscripts by early Latin letters, namely $a,b,c=1,\bar1,2,\bar2$, and the components belonging to the Lie algebra of the subgroup $H$ in $G=SU(3)$ will be denoted by $i,j,k,l=5,...,8$. The letters $d,e$ denote only real coset indices, here either $1$ or $2$.

As a matter of fact, the Hermitian metric with respect to this basis has components only with mixed holomorphic and anti-holomorphic indices and therefore we get
\begin{equation}\label{invariant metric cp2}
 \dr s_{G/H}^2 = \de_{d \bar e}~\beta^{d} \bar\beta^{\bar e}
\end{equation}
Pulling down indices with this particular metric, we obtain that indices get complex conjugated
\begin{equation*}
 T^a=T_{\bar a}~\de^{\bar aa}\,.
\end{equation*}
Our metric on the product space becomes
\begin{equation*}
 \dr s^2 = \dr \tau^2 + \de_{d \bar e}~\beta^{d} \bar\beta^{\bar e}\,,
\end{equation*}
which allows us to pull down $x^0=\tau$ indices without changing the coefficients:
\begin{equation*}
 T^0 = T_0 ~\de^{00}\,,~\text{with}~\de^{00}=1\,.
\end{equation*}

%\subsection{The Generators of $G/H$}

\paragraph{The symmetric \boldmath{$\underline{C}^{1,0}$} quiver bundle.}
\label{the symmetric C10 quiver bundle}
We want to start with the simplest case of a quiver theory for $\BC P^2$ in order to see how it works. We will not use this specific ansatz to derive Yang-Mills equations, since considerations with one scalar field were already done in \cite{yangmillsflow} and would lead to similar results here.
The space we are dealing with is given by the quotient
\begin{equation*}
 \BC P^2 = \frac{SU(3)}{SU(2)\times U(1)}\,,
\end{equation*}
which is a symmetric space.

One can employ Young tableaux in order to get the decomposition of the fundamental representation of $SU(3)$ into irreducible representations of $H$. We have
\begin{equation}
\underline{C}^{1,0}\big|_{SU(2)\times U(1)}=\underline{(1,1)}~\oplus~
\underline{(0,-2)} \,,
\label{decomposition1CP1}
\end{equation}
where in $\underline{(n,m)}$ we mean $n=2I$ to take two times the values of the isospin $I$, namely the eigenvalues of the first generator of the Cartan subalgebra (denoted by $E_7$ in (\ref{Hgenerators}) below), such that the dimension of the corresponding irreducible representation equals $n+1$. So the first piece of the sum in (\ref{decomposition1CP1}) is going to be two-dimensional while the second piece is going to be one-dimensional. The second number in the brackets, $m=3Y_H$, is meant to equal three times the hypercharge $Y_H$, which can be associated to the corresponding eigenvalues of the second generator of the Cartan subalgebra (denoted by $E_8$ in (\ref{Hgenerators}) below).
From (\ref{decomposition1CP1}) we can already see that there may be only one arrow between the irreps of $H=SU(2)\times U(1)$ and hence we are getting one scalar field in our gauge potential. We can easily see that the structure group in this example equals $U(3)$. The quiver diagram for this case is
\begin{equation*}
\xymatrix{ & \BR\otimes \CV_{(1,1)} \\
	  \ar[ur]^{\phi\otimes\beta}\BR\otimes \CV_{(0,-2)} & }
\end{equation*}

In the following we are going to write down the $SU(3)$-equivariant connection for the corresponding vector bundle explicitly which requires the explicit form of the generators. For the fundamental 3-dimensional representation of $SU(3)$ the generators corresponding to $G/H=\mathbb CP^2$ are given by
\begin{equation}
\begin{aligned}
 &E_1=e_{31}\,,& \quad &E_2=e_{32}\,,& \quad &E_{\bar1}=e_{13}\,,& \quad &E_{\bar2}=e_{23}\,,&\\
\end{aligned}
\end{equation}
and the generators of $H=SU(2)\times U(1)$ are given by
\begin{equation}
\begin{aligned}
 &E_5=e_{12}\,,& \quad &E_6=e_{21}\,,& \quad &E_7=e_{11}-e_{22}\,,& \quad &E_8=e_{11}+e_{22}-2e_{33}\,,& \label{Hgenerators}
\end{aligned}
\end{equation}
where we used the notation of the matrix units for $3\times3$ matrices, defined by
\begin{equation*}
 \left(e_{ij} \right)_{kl} =\de_{ik}\de_{jl}\,.
\end{equation*}

Let us consider a flat connection on the trivial bundle $\BC P^2\times \BC^3$ over $\BC P^2$, given by
\begin{equation}\label{ansatz cp2 c10 flat connection}
A_0=~\begin{pmatrix} B&\bar\beta\\[4pt]-\beta^\top&-2a
\end{pmatrix}\,,
\end{equation}
where
\begin{subequations}
\begin{eqnarray}
 B&=&\frac1{\gamma^2}\,\big(-\mbox{$\frac12$}~\dr(Y^\dag\,Y)~
\idm_2+\bar Y~
\dr\bar Y^\dag+\Lambda~\dr\Lambda\big) \,, \\
a&=&-\frac1{4\gamma^2}\,\big(\bar Y^\dag~\dr\bar Y-\dr\bar Y^\dag~\bar Y\big) \,,
\end{eqnarray}
\end{subequations}
along with the notation from (\ref{gammaYdef}) as well as
\begin{equation*}
\Lambda~:=~\gamma~\idm_2-\frac1{\gamma+1}~Y\,Y^\dag
\end{equation*}
It satisfies the Maurer-Cartan equation which reads
\begin{equation}
 \dr A_0+A_0\wedge A_0=0 \,,
\end{equation}
yielding
\begin{subequations}
\begin{eqnarray}
 \label{maurcart-dB}
 \dr B + B \wedge B - \bb \wedge \be^\top &=& 0\,,\\
\label{maurcart-da}
 \dr a - \frac{1}{2} \be^\dagger\wedge \be &=& 0\,,\\
\label{maurcart-dbetabar}
 \dr \bb + B\wedge \bb -2\bb\wedge a &=& 0\,,\\
\label{maurcart-dbetatransp}
 \dr \beta^\top+\beta^\top\wedge B-2a\wedge\beta^\top & = & 0\,.
\end{eqnarray}
\end{subequations}

Using these formulae, one can extend the flat connection on the trivial bundle over $\BC P^2$ to a connection on the bundle over $\BR\times \BC P^2$. It is given by the $3\times 3$ matrix:
\begin{equation}\label{ansatz1b}
 \CA = \left(\begin{matrix}
      B_{(1)}+a~\idm_2 & \phi \bb\\ -\phi \be^\top & -2a
     \end{matrix}\right)\,,
\end{equation}
where we identify the $su(2)$-valued one-instanton field $B_{(1)}$ on $\BC P^2$ with the $2\times 2$ matrix
\begin{equation}\label{def B_(1)}
 B_{(1)}:=B-a\idm_2=:\left(\begin{matrix}B^{11} & B^{12}\\ -\overline{B^{12}} & -B^{11}\end{matrix}\right)\,.
\end{equation}
The corresponding field strength is easily calculated using (\ref{maurcart-dB})-(\ref{maurcart-dbetatransp}) and takes the form
\begin{equation}
\CF = \left(\begin{matrix}
       (1-\phi^2) ~ (\bb\wedge\be^\top) & \dot\phi ~ \dr \tau \wedge \bb \\
	-\dot\phi ~ \dr\tau\wedge \be^\top  & -(1-\phi^2) ~ (\be^\dagger \wedge \be)
      \end{matrix}\right)\,,
\end{equation}
with
\begin{eqnarray*}
 \left(\begin{matrix}
	\bar\beta^1\wedge\beta^1 & \bar\beta^1\wedge\beta^2\\
	\bar\beta^2\wedge \beta^1& \bar\beta^2\wedge \beta^2
  \end{matrix}\right)
&=& \bar\beta \wedge \beta^\top \,,\\
(\bar\beta^1\wedge\beta^1+\bar\beta^2\wedge\beta^2)
&=& \beta^\dagger\wedge \beta\,,
\end{eqnarray*}
and
\begin{equation*}
 \dot\phi=\frac{\dr\phi(\tau)}{\dr\tau}\,.
\end{equation*}

Using the explicit form of the generators in the fundamental representation of $SU(3)$, we can write the Maurer-Cartan form as
\begin{eqnarray}
 A_0 &=& -\be^1~E_1 - \be^2~E_2 + \bb^{\bar1}~E_{\bar1} + \bb^{\bar2}~E_{\bar2} %\nonumber\\
      + B^{12}~E_5 - \overline{B^{12}}~E_6 + B^{11}~E_7 + a~E_8
\label{maurer cartan form cp2 general}
\end{eqnarray}
and hence (\ref{ansatz1b}) is nothing but
\begin{eqnarray}\label{gauge potential form cp2 c10 general}
 \CA &=& \phi \left(-\be^1 E_1 - \be^2 E_2 + \bb^{\bar1} E_{\bar1} + \bb^{\bar2} E_{\bar2}\right) +  e^i{}_b E_i~e^b\,,
\end{eqnarray}
where
\begin{eqnarray}\label{e^i_b cp2}
\begin{aligned}
 &e^1=\be^1,& &e^2=\be^2,& &e^{\bar1}=\bb^{\bar1},& &e^{\bar2}=\bb^{\bar2}\,,&\\
 &e^5{}_b = B^{12}_b,& &e^6{}_b=-\overline{B^{12}_b},& &e^7{}_b=B^{11}_b,& &e^8{}_b=a_b\,.&
\end{aligned}
\end{eqnarray}
We will need (\ref{e^i_b cp2}) later on in order to differentiate the field strength covariantly. From (\ref{gauge potential form cp2 c10 general}) we can see that this ansatz would yield the same results we already deduced in \cite{yangmillsflow}. Getting more scalar fields involved requires the choice of a higher dimensional representations of $SU(3)$ which we will do in the following.
\paragraph{The symmetric \boldmath{$\underline{C}^{2,0}$} quiver bundle.}
%
%
%Since we have already solved the problem for the ansatz of a 3-dimensional quiver representation,
For $\BC P^2$ we have seen so far how the quiver bundle looks for the case of fundamental representation of $SU(3)$. We now want to use the generalizations to the 6-dimensional representation $\underline{C}^{2,0}$ of $SU(3)$. An important point is that (\ref{maurer cartan form cp2 general}) actually holds for arbitrary quiver representations by inserting the corresponding higher dimensional generators. Specifically for $\underline{C}^{2,0}$, we have the following generators:
\begin{eqnarray}\label{HgeneratorsC20}
\begin{aligned}
&E_{1}=\sqrt2\,\big(e_{41}+e_{64}\big)+e_{52}\,,&\quad
&E_{\bar1}=\sqrt2\,\big(e_{14}+e_{46}\big)+e_{25}\,,&\\
&E_{2}=e_{42}+\sqrt2\,\big(e_{53}+e_{65}\big)\,,&\quad
&E_{\bar2}=e_{24}+\sqrt2\,\big(e_{35}+e_{56}\big)\,,&\\
&E_{5}=\sqrt2\,\big(e_{12}+e_{23}\big)+e_{45}\,,&\quad
&E_{6}=\sqrt2\,\big(e_{21}+e_{32}\big)+e_{54}\,,&\\
&E_{7}=2(e_1-e_3)+e_4-e_5\,,&\quad
&E_{8}=2(e_1+e_2+e_3)-e_4-e_5-4e_6\,.&
\end{aligned}
\end{eqnarray}

Using the formalism of Young tableaux again, we find the following decomposition into irreducible subspaces:
\begin{equation}
\underline{C}^{2,0}\big|_{SU(2)\times U(1)}=\underline{(2,2)}~\oplus~
\underline{(1,-1)}~\oplus~\underline{(0,-4)}\,.
\end{equation}
The corresponding quiver diagram is then given by
\begin{equation*}
\xymatrix{ & & \BR\otimes \CV_{(2,2)} \\
	   &\ar[ur]^{\phi_1\otimes \beta_1}\BR\otimes \CV_{(1,-1)} &\\
	    \ar[ur]^{\phi_2\otimes \beta_2}\BR\otimes \CV_{(0,-4)} & & }
\end{equation*}
\paragraph{Gauge potential and field strength.}
As we can see, one ends up with an $SU(3)$-equivariant connection containing two Higgs fields $\phi_1(\tau),~\phi_2(\tau)$ which is a connection on the corresponding associated vector bundle (\ref{general associated principal quiver bundle}) with the structure group $U(6)$. This gauge potential is in general given in \cite{preprint5} and in our case it simplifies to the block $6\times6$ matrix 
\begin{equation}\label{ansatz gauge potential cp2c20}
 \CA := \left(\begin{matrix}
		B_{(2)} + 2~a~\idm_3 & \phi_1 \bar\beta_1 & 0 \\
		-\phi_1 \bar\beta_1^\dagger & B_{(1)}-a~\idm_2 & \phi_2 \bar\beta_2 \\
		0 & -\phi_2 \bar\beta_2^\dagger & -4~a
           \end{matrix}\right)
\end{equation}
where the one-instanton connection $B_{(2)}$ in the 3-dimensional irreducible representation of $SU(2)$ is defined as
\begin{equation}\label{def B_(2)}
 B_{(2)}=\begin{pmatrix} 2B^{11}&\sqrt2\,B^{12}&0\\[4pt]
-\sqrt2~\overline{B^{12}}&0&\sqrt2\,B^{12}\\[4pt]
 0&-\sqrt2~\overline{B^{12}}&-2B^{11}\end{pmatrix}\,.
\end{equation}
The matrices $\bar\beta_1$ and $\bar\beta_2$
%(not to be confused with the invariant one-forms $\bar\beta^{\bar1}$ and $\bar\beta^{\bar1}$)
are given by
\begin{equation*}
 \bb_1=\begin{pmatrix}\sqrt2~\bb^{\bar1}&0\\[4pt]
\bb^{\bar2}&\bb^{\bar1}\\[4pt]0&\sqrt2~\bb^{\bar2}\end{pmatrix}
\qquad \mbox{and} \qquad \bb_2=\sqrt2\,
\begin{pmatrix}\bb^{\bar1}\\[4pt]\bb^{\bar2}\end{pmatrix} \,.
\label{betaC20rep}
\end{equation*}

We also take the field strength from \cite{preprint5}, which is obtained from $\CF=\dr \CA + \CA\wedge \CA$ using the fact that the flat connection (\ref{maurer cartan form cp2 general}) satisfies the Maurer-Cartan equations. One arrives at the following field strength:
\begin{equation}
 \CF = \begin{pmatrix}
        (1-\phi_1^2)~\bb_1\wedge\bb_1^\dagger & \dot\phi_1~ \dr\tau\wedge \bb_1 & 0 \\
	 -\dot\phi_1~\dr\tau\wedge \bb_1^\dagger & \begin{matrix} (1-\phi_1^2)~\bb_1^\dagger\wedge\bb_1 \\
	 +(1-\phi_2^2)~\bb_2\wedge\bb_2^\dagger\end{matrix} & \dot\phi_2~\dr \tau\wedge \bb_2\\
	0& -\dot\phi_2~\dr \tau\wedge \bb_2^\dagger & (1-\phi_2^2)~\bb_2^\dagger\wedge\bb_2
       \end{pmatrix}\,,
\end{equation}
where
\begin{eqnarray*}
\bb_1\wedge\bb_1^\dag & = & 
\begin{pmatrix} 2\,\bb^1\wedge\be^1&\sqrt2~\bb^1\wedge\be^2&0\\[4pt]
\sqrt2~\bb^2\wedge\be^1&\bb^1\wedge\be^1+\bb^2\wedge
\be^2&\sqrt2~\bb^1\wedge\be^2\\[4pt]
0&\sqrt2~\bb^2\wedge\be^1&2\,\bb^2\wedge\be^2 \end{pmatrix} \ ,
\label{beta11prod1CP2C20}\\[4pt]
\bb_1^\dag \wedge\bb_1 &=&
-\begin{pmatrix}2\,\bb^1\wedge\be^1+\bb^2\wedge
\be^2&\bb^1\wedge\be^2\\[4pt]
\bb^2\wedge\be^1&\bb^1\wedge\be^1+2\,\bb^2\wedge\be^2 \end{pmatrix} \ ,
\label{beta11prod2CP2C20}\\[4pt]
\bb_2\wedge\bb_2^\dag&=&2\,
\begin{pmatrix}\bb^1\wedge\be^1&\bb^1\wedge\be^2\\[4pt]
\bb^2\wedge\be^1&\bb^2\wedge\be^2 \end{pmatrix} \ ,
\label{beta04prod1CP2C20}\\[4pt]
\bb_2^\dag\wedge\bb_2&=&-2\,
\big(\bb^1\wedge\be^1+\bb^2\wedge\be^2\big) \ .
\label{beta04prod2CP2C20}
\end{eqnarray*}
\paragraph{Yang-Mills equations.}
\label{ansatz c11 cp2 section yang mills}
%Now we collected all the information needed to write down the Yang-Mills equations and are prepared to 
Now we shall derive the corresponding differential equations for the scalar fields $\phi_1$ and $\phi_2$. The Levi-Civita connection 1-form on $\BC P^2$ for the invariant metric (\ref{invariant metric cp2}) is given by formulae
\begin{eqnarray}\label{connection 1-form cp2}
 \omega^{c}{}_b = f_{ib}{}^c e^i = -f_{bi}{}^c e^i,\quad e^i=e^i{}_a e^a
\end{eqnarray}
with $e^i{}_a$ from (\ref{e^i_b cp2}). Here, we also used the fact that $\BC P^2$ is a symmetric space and hence
\begin{equation*}
 f_{ab}{}^{c} = 0\quad \forall a,b,c\in \left\{1,2,\bar1,\bar2\right\}\,.
\end{equation*}
We easily find the following non-vanishing structure constants $f_{bi}{}^c$ of $\BC P^2$:
\begin{eqnarray}\label{structure constants cp2}
 \begin{aligned}
  f_{15}{}^2&=&1 \qquad  f_{\bar16}{}^{\bar2}&=&-1 \qquad f_{26}{}^1&=&1 \qquad  f_{\bar25}{}^{\bar1}&=&-1\\
  f_{17}{}^1&=&1 \qquad  f_{\bar17}{}^{\bar1}&=&-1 \qquad f_{27}{}^2&=&-1\qquad f_{\bar27}{}^{\bar2}&=&1\\
  f_{18}{}^1&=&3 \qquad  f_{\bar18}{}^{\bar1}&=&-3 \qquad f_{28}{}^2&=&3 \qquad f_{\bar28}{}^{\bar2}&=&-3 \,.\\
  \end{aligned}
\end{eqnarray}
Clearly, these are nothing but the structure constants of $SU(3)$ in the basis $E_A,~A\in\{1,...,8\}$ introduced in (\ref{HgeneratorsC20}).
Since we use the direct product metric on $\BR \times \BC P^2$, we have
\begin{eqnarray}
 \omega^0_{0b}=\omega^a_{0b}=\omega^0_{cb} = 0\,, 
\end{eqnarray}
and the non-vanishing components are
\begin{eqnarray}\label{ansatz cp2 c20 connection 1 form}
 \omega^c_{ab}=-f_{bi}{}^c~e^i{}_a~e^a\,.
\end{eqnarray}
So, the Yang-Mills equations read
\begin{eqnarray}
 \CD_a \CF^{a0} = 0\,, \label{YM cp2 c20 1}\\
 \CD_0 \CF^{0b} + \CD_a \CF^{ab} = 0\,, \label{YM cp2 c20 2}
\end{eqnarray}
where $\CD_0:=\frac{\dr}{\dr \tau}$ and $\CD_a \CF^{ab}:=\e_a(\CF^{ab}) + \omega^a_{ac} \CF^{cb} + \omega^b_{ac} \CF^{ac}+ \com{\CA_a}{\CF^{ab}}$.

In order to simplify these equations, we use the splitting of the gauge potential in its block-diagonal and off-diagonal parts (\ref{gauge potential splitting}). Inserting this splitting of the gauge potential and (\ref{ansatz cp2 c20 connection 1 form}) into (\ref{YM cp2 c20 1}) and (\ref{YM cp2 c20 2}), we get
\begin{eqnarray}
\label{ansatz cp1 c20 yang mills equation1}
 0&=& \overbrace{-e^i{}_a\left(f_{ci}{}^a~\CF^{c0}\right)+\left[\CA_a^{\text{diag}},\CF^{a0}\right] + \left[\CA_a^{\text{off}},\CF^{a0}\right]}^{=~0,~\textrm{trivially}}\,,\\
 0&=&\frac{\dr}{\dr\tau} \CF^{0b}-e^i{}_a\left(f_{ci}{}^a~\CF^{cb}+ f_{ci}{}^b~\CF^{ac} \right)+\left[\CA_a^{\text{diag}},\CF^{ab}\right] + \left[\CA_a^{\text{off}},\CF^{ab}\right] \,.
\label{ansatz cp1 c20 yang mills equation2}
\end{eqnarray}
We find that equation (\ref{ansatz cp1 c20 yang mills equation1}) is trivially satisfied and therefore yields no restrictions on the fields. From equation (\ref{ansatz cp1 c20 yang mills equation2}) we get
\begin{equation}\label{ansatz cp1 c20 yang mills equations vanishing piece}
 e^i{}_a\left(f_{ci}{}^a~\CF^{cb}+ f_{ci}{}^b~\CF^{ac} \right) = \left[\CA_a^{\text{diag}},\CF^{ab}\right]\,,
\end{equation}
and therefore (\ref{ansatz cp1 c20 yang mills equation2}) becomes
\begin{equation}\label{ansatz cp1 c20 yang mills equations non vanishing piece}
  0=\frac{\dr}{\dr\tau} \CF^{0b} + \left[\CA_a^{\text{off}},\CF^{ab}\right]\,,
\end{equation}
which for every index $b$ leads to a matrix equation containing two independent differential equations:
%
% \begin{eqnarray}
%   \quad \ddot \phi_1 &=& -\phi_1~\left(3 - (5\phi_1^2 - 2\phi_2^2)\right) \label{ansatz cp2 c20 eom 1}\\
%   \quad \ddot \phi_2 &=& -\phi_2~\left(3 - (6\phi_2^2 - 3\phi_1^2) \right) \label{ansatz cp2 c20 eom 2}
% \end{eqnarray}
\begin{subequations}
\begin{eqnarray}
  \quad \ddot \phi_1 &=& 3\phi_1~\left( \frac53\phi_1^2 -1 - \frac23\phi_2^2\right) \label{ansatz cp2 c20 eom 1}\\
  \quad \ddot \phi_2 &=& 3\phi_2~\left( 2\phi_2^2 -1 -\phi_1^2 \right) \label{ansatz cp2 c20 eom 2}
\end{eqnarray}
 \label{ansatz cp2 c20 eom}
\end{subequations}

Here, we can already recognize that for $\phi_1=\phi_2=\phi$ we obtain only one differential equation, similar that from \cite{yangmillsflow}, namely
\begin{equation}\label{ansatz cp2 c20 eom 3}
 \ddot \phi = 3\phi~\left(\phi^2-1\right)\,,
\end{equation}
which is solved for instance by
\begin{equation}\label{ansatz cp2 c20 solution 1}
 \phi(\tau) = \tanh{\left(\sqrt{\frac32}~\tau\right)}\,.
\end{equation}

If we put one of the $\phi_i$ to zero we get either
\begin{eqnarray}\label{ansatz cp2 c20 eom 4}
 \ddot \phi_1 &=& -\phi_1~\left(3 - 5\phi_1^2\right)\,,\qquad \phi_2=0
\end{eqnarray}
or
\begin{equation}\label{ansatz cp2 c20 eom 5}
\phi_1=0\,,\qquad \ddot \phi_2 = -\phi_2~\left(3 - 6\phi_2^2 \right)\,.
\end{equation}
These two equations can also be solved by a hyperbolic tangens, for instance
\begin{eqnarray}
 \phi_1(\tau) = \sqrt{\frac35} \tanh{\left(\sqrt{\frac32}~\tau\right)}\,, \qquad \phi_2=0\,,
\end{eqnarray}
and
\begin{eqnarray}
 \phi_1=0 \,, \qquad \phi_2(\tau) = \sqrt{\frac12} \tanh{\left(\sqrt{\frac32}~\tau\right)}\,,
\end{eqnarray}
% ------------------------
% \begin{equation}
% \phi_1 = 0\,,\qquad \phi_2 \left( \tau \right) =\sqrt {1-{C_1}^{2}}~\text{sn} \left( \sqrt {3}~C_1~\tau+
% C_2,\sqrt {{\frac {1-{C_1}^{2}}{{C_1}^{2}}}} \right)
% \,,
% \end{equation}
% ---------------------
respectively.% where $C_1$ and $C_2$ are constants. In the first case we chose the constants such that one obtains a hyperbolic tangens. Since we need the second argument of the Jacobi elliptic function $\text{sn}\left(\cdot,\cdot\right)$ to be 1 in order to obtain such a hyperbolic tangens, this cannot be achieved for the solution $\phi_2$ in the case where $\phi_1$ vanishes.
\begin{figure}[ht]
\subfigure[Solutions  of (\ref{ansatz cp2 c20 eom}) with the following initial values:\newline $\phi_1(0)=0,~\dot\phi_1(0)=1,~\phi_2(0)=0,~\dot\phi_2(0)=0.6$]{
  \includegraphics[width=7.9cm]{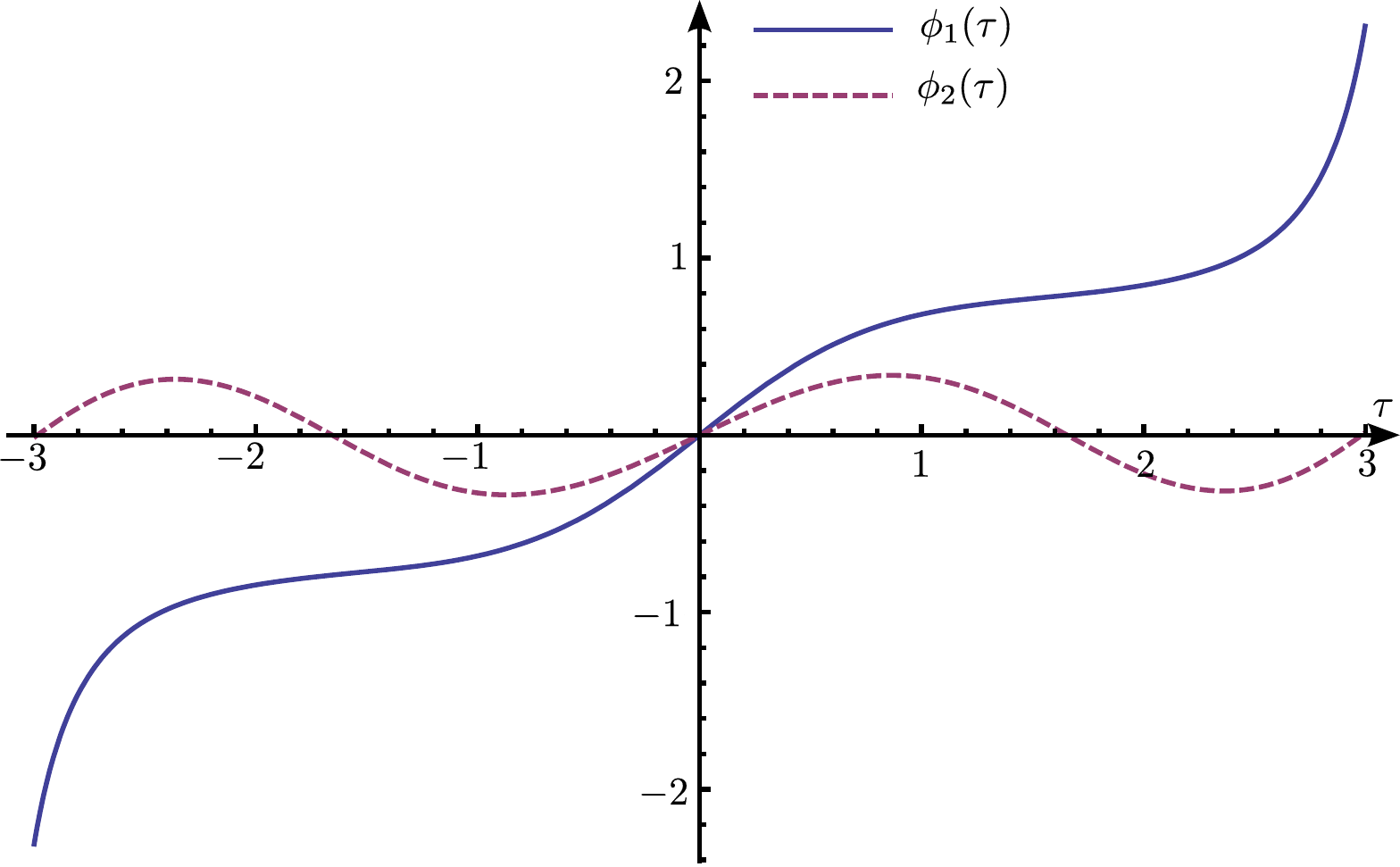}
}
\subfigure[Solutions of (\ref{ansatz cp2 c20 eom}) with the following initial values:\newline
$\phi_1(0)=1,~,\dot\phi_1(0)=0,~\phi_2(0)=0.6,~\dot\phi_2(0)=0$]{
 \includegraphics[width=7.9cm]{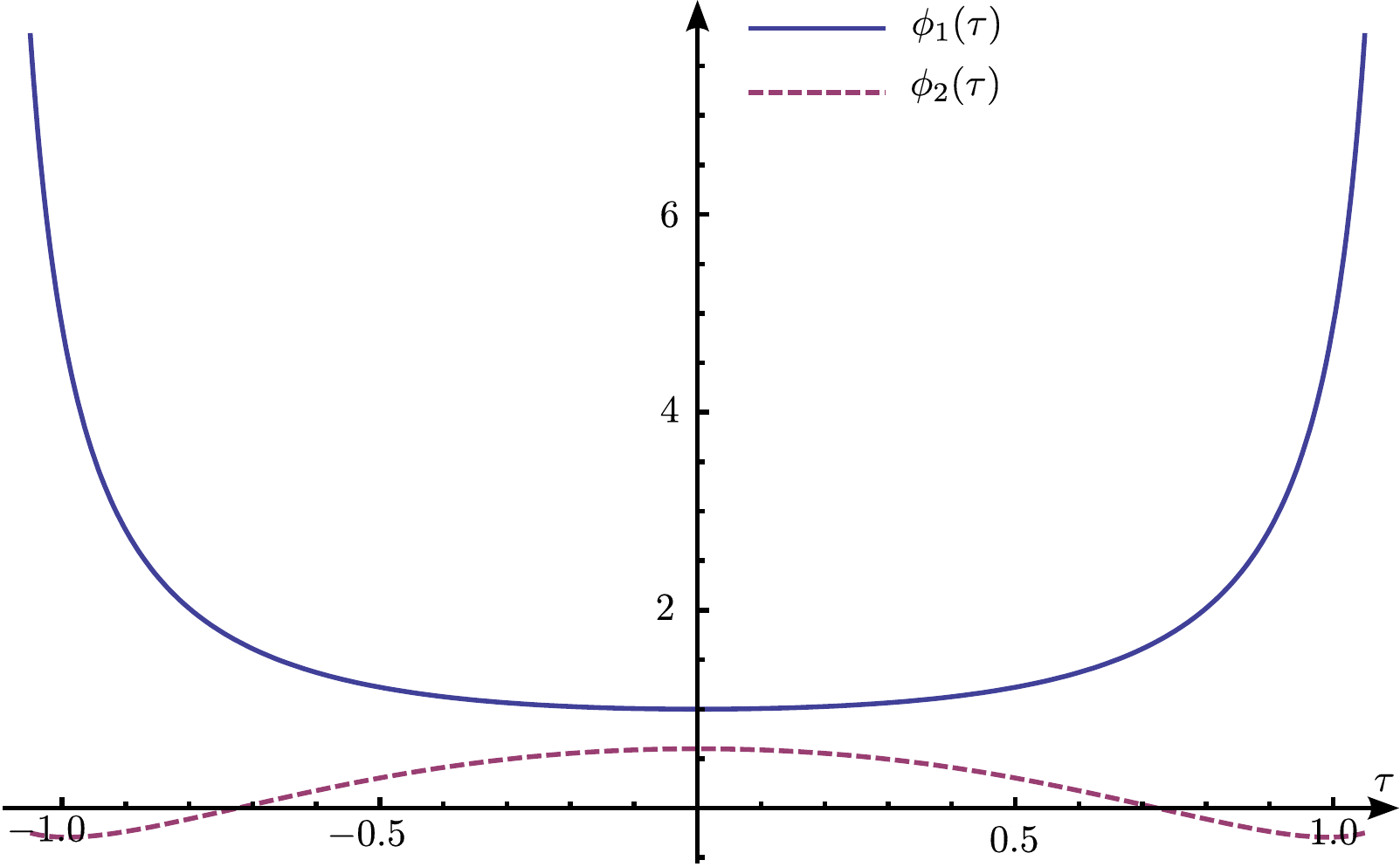}
}
 \caption{\small Two different sets of truly coupled numerical solutions $\phi_1(\tau),~\phi_2(\tau)$ of the system of differential equations (\ref{ansatz cp2 c20 eom}). Particularly $\phi_1\neq\phi_2,~\phi_1\neq0\neq\phi_2$.}
  \label{fig:RxCP2-C20}
\end{figure}
What remains is the task of finding ``truly coupled'' solutions to the system (\ref{ansatz cp2 c20 eom}). In general we refer to  truly coupled solutions of a system of differential equations as of a set of fields $\{\phi_1,...,\phi_k\}$ solving the system and satisfying:
\begin{equation*}
 \forall~i\in\{1,...,k\}~\exists~\tau, \text{ such that } \phi_i(\tau)\neq0 \text{ and } \forall~i,j\in \{1,...,k\}~\exists~\tau \text{ such that } \phi_i(\tau)\neq\phi_j(\tau)\,,
\end{equation*}
which simply means that no two of fields in the solution set coincide and none of the fields is trivial. Finding the most general exact truly coupled solutions to the equations of motion (\ref{ansatz cp2 c20 eom}) is not an easy task and we will not provide them here, but at least by making use of a computer, we found some numerical solutions to this problem for certain specific initial values which are given in figure \ref{fig:RxCP2-C20}.
\section[Yang-Mills theory on R x CP2 in C(1,1) quiver representation]{Yang-Mills theory on $\mathbb R \times \BC P^2$ in $\underline{C}^{1,1}$ quiver representation}
%\section{YANG-MILLS THEORY ON $\mathbb R \times \BC P^2$ IN $\underline{C}^{1,1}$ QUIVER REPRESENTATION}
\addcontentsline{toc}{section}{Yang-Mills on R x CP2 in C(1,1) quiver representation}
 We now go one step further and choose a representation that decomposes into more irreps in order to obtain more than two scalar fields. So we are going to take the gauge connection and field strength from \cite{preprint5} and use them for deriving the corresponding Yang-Mills equations and the equations of motion for the Higgs fields explicitly.
\paragraph{The symmetric \boldmath{$\underline{C}^{1,1}$} quiver bundle.}
We are choosing the $\underline{C}^{1,1}$ highest weight representation of $SU(3)$ which is eight dimensional and therefore its adjoint representation. We obtain the following decomposition after the restriction to $H=SU(2)\times U(1)$:
\begin{equation}
\underline{C}^{1,1}\big|_{SU(2)\times U(1)}=\underline{(1,-3)}~\oplus~
\underline{(2,0)}~\oplus~\underline{(0,0)}~\oplus~\underline{(1,3)} \,.
\end{equation}
We have the following quiver diagram:
\begin{equation*}
\xymatrix{ & \BR\otimes \CV_{(1,3)} &  \\
	   \ar[ur]^{\phi_3\otimes\beta_3}\BR\otimes \CV_{(0,0)}& &\ar[ul]_{\phi_4\otimes\beta_4}\BR\otimes \CV_{(2,0)}\\
	    &\ar[ul]^{\phi_1\otimes\beta_1}\ar[ur]_{\phi_2\otimes\beta_2}\BR\otimes \CV_{(1,-3)}&}
\end{equation*}
From this one can already see that there will appear four independent scalar fields in the gauge connection.
\paragraph{Gauge potential and field strength.}
The generators of this eight-dimensional representation can be written in terms of the eight-dimensional matrix units:
\begin{align*}
  &E_1=e_{12}+\sqrt{2}~(e_{45}+e_{56})+e_{78}\,,&\\
  &E_{\bar1}=e_{21}+\sqrt{2}~(e_{54}+e_{65})+e_{87}\,,&\displaybreak[0]\\
  &E_2=e_{14}+\sqrt{\frac32}~(e_{23}+e_{37})+\sqrt{\frac12}(e_{25}+e_{57})+e_{68}\,,&\\ &E_{\bar2}=e_{41}+\sqrt{\frac23}~(e_{32}+e_{73})+\sqrt{\frac12}(e_{52}+e_{75})+e_{86}\,,&\displaybreak[0]\\
  &E_5=\sqrt{\frac32}(e_{13}-e_{38} ) - \sqrt{\frac32}(e_{15}-e_{58})+(e_{47} - e_{26})\,,&\\
  &E_6=\sqrt{\frac23}(e_{31}-e_{83} ) - \sqrt{\frac23}(e_{51}-e_{85})+(e_{74} - e_{62})\,,&\displaybreak[0]\\
  &E_7= (e_{11}-e_{22} )+2(e_{44}-e_{66})+(e_{77}-e_{88})\,,&\\
  &E_8=3(e_{11}+e_{22}-e_{77}-e_{88})\,,&
\end{align*}
where as before the generators with subscripts $\left\{1,2,\bar1,\bar2\right\}$ correspond to the coset space and the others with subscripts $\left\{5,6,7,8\right\}$ denote the generators of the subgroup $H$.

We also see from the decomposition with respect to the subgroup $H$ that the associated vector bundle given in (\ref{general structure group}) comes with the structure group $U(8)$ in this case. The corresponding $SU(3)$-equivariant connection is then given (equation (3.125) in \cite{preprint5}) by
\begin{equation}\label{ansatz c11 cp2 gauge potential}
 \CA = \left(\begin{matrix}
              B_{(1)}+3a~\idm_2 & \phi_3~\bb_3 & \phi_4~\bb_4 & 0 \\
	      -\phi_3~\bb_3^\dagger & 0 & 0 & \phi_1~\bb_1 \\
	      -\phi_4~\bb_4^\dagger & 0 & B_{(2)} & \phi_2~\bb_2\\
	      0 & -\phi_1~\bb_1^\dagger & -\phi_2~\bb_2^\dagger & B_{(1)}-3a~\idm_2
              \end{matrix}\right),
\end{equation}
with $B_{(1)}$, $B_{(2)}$ from (\ref{def B_(1)}), (\ref{def B_(2)}) and
\begin{equation}\label{beta_i matrices c11 cp2}
\begin{aligned}
\bb_3=\sqrt{\frac32}~\begin{pmatrix}\bb^{\bar{1}}\\
\bb^{\bar{2}}\end{pmatrix}, \qquad & & \qquad
\bb_4=\begin{pmatrix}\bb^{\bar{2}}&-\sqrt{\frac12}~
\bb^{\bar{1}}&0\\0&\sqrt{\frac12}~\bb^{\bar{2}}&-\bb^{\bar{1}}
\end{pmatrix} \ , \\
\bb_1=\sqrt{\frac32}~\big(\bb^{\bar{2}}\,,\,-
\bb^{\bar{1}}\big), \qquad & & \qquad
\bb_2=\begin{pmatrix}\bb^{\bar{1}}&0\\
\sqrt{\frac12}~\bb^{\bar{2}}&\sqrt{\frac12}~\bb^{\bar{1}}
\\ 0&\bb^{\bar{2}}\end{pmatrix} \ .
 \end{aligned}
\end{equation}
We have the following field strength
\begin{small}
\begin{equation}\label{ansatz c11 cp2 gauge field strength}
\begin{aligned}
 &\CF=\dr \CA +\CA\wedge \CA = 
\\&\left(\begin{matrix}
      \begin{matrix}(1-\phi_3^2)~\bb_3\wedge\bb_3{}^\dagger\\
	+(1-\phi_4^2)~\bb_4\wedge\bb_4{}^\dagger \end{matrix}
		&\hspace{-0.6cm} \dr\phi_3\wedge\bb_3
		&\hspace{-0.3cm} \dr\phi_4\wedge\bb_4
		&\hspace{-0.5cm} (\phi_3\phi_1-\phi_4\phi_2)~\bb_3\wedge\bb_1 \\[0.3cm] %\hline
     -\dr\phi_3\wedge\bb_3{}^\dagger
		&\hspace{-0.6cm} (\phi_3^3-\phi_1^3)~\bb_1\wedge\bb_1^\dagger
		&\hspace{-0.3cm} (\phi_3\phi_4-\phi_1\phi_2)~\bb_1\wedge\bb_2{}^\dagger
		&\hspace{-0.5cm} \dr\phi_1\wedge\bb_1 \\[0.3cm] %\hline
     -\dr\phi_4\wedge\bb_4{}^\dagger
		&\hspace{-0.6cm} (\phi_1\phi_2-\phi_3\phi_4)~\bb_1{}^\dagger\wedge\bb_2
		&\hspace{-0.3cm} \begin{matrix}(1-\phi_4^2)~\bb_4{}^\dagger\wedge\bb_4 \\ +(1-\phi_2^2)~\bb_2\wedge\bb_2{}^\dagger \end{matrix}
		&\hspace{-0.5cm} \dr\phi_2\wedge\bb_2 \\[0.3cm] %\hline
      (\phi_4\phi_2-\phi_3\phi_1)~\bb_3{}^\dagger\wedge\bb_1{}^\dagger
		&\hspace{-0.6cm} -\dr\phi_1\wedge\bb_1{}^\dagger
		&\hspace{-0.3cm} -\dr\phi_2\wedge\bb_2{}^\dagger
		&\hspace{-0.5cm} \begin{matrix} (1-\phi_1^2)~\bb_1^\dagger\wedge\bb_1\\+(1-\phi_2^2)~\bb_2^\dagger\wedge\bb_2\end{matrix}
     \end{matrix} \right)\\[4pt]
\end{aligned}
\end{equation}
\end{small}
The wedge product expressions of the $\beta_i$ matrices from (\ref{beta_i matrices c11 cp2}) are given by
\begin{align*}
\bb_3\wedge\bb_3{}^\dag
=&~\frac32\,\begin{pmatrix}
\bb^{\bar1}\wedge\beta^1&\bb^{\bar1}\wedge\beta^2\\\bb^{\bar2}\wedge\beta^1&
\bb^{\bar2}\wedge\beta^2\end{pmatrix} \ , 
%\label{adbetap00}
\displaybreak[0]\\[4pt]
\bb_4\wedge\bb_4{}^\dag
=&~\begin{pmatrix}
\frac12\,\bb^{\bar1}\wedge\beta^1+\bb^{\bar2}\wedge\beta^2&
-\frac12\,\bb^{\bar1}\wedge\beta^2\\-\frac12\,\bb^{\bar2}\wedge\beta^1&
\bb^{\bar1}\wedge\beta^1+\frac12\,\bb^{\bar2}\wedge\beta^2\end{pmatrix} \ ,
%\label{adbetam20}
\displaybreak[0]\\[4pt]
\bb_4{}^\dag\wedge\bb_4
=&~\begin{pmatrix}
\beta^2\wedge\bb^{\bar2}&-\sqrt{\frac12}~\beta^2\wedge\bb^{\bar1}&0\\
-\sqrt{\frac12}~\beta^1\wedge\bb^{\bar2}&\frac12\,\big(\beta^1\wedge
\bb^{\bar1}+\beta^2\wedge
\bb^{\bar2}\big)&-\sqrt{\frac12}~\beta^2\wedge\bb^{\bar1}\\
0&-\sqrt{\frac12}~\beta^1\wedge\bb^{\bar2}&\beta^1\wedge\bb^{\bar1}
\end{pmatrix} \ , 
%\label{adbetam20dag}
\\[4pt]
 \displaybreak[0]
\bb_1\wedge\bb_1{}^\dag
=&~\mbox{$\frac32$}\,\big(
\bb^{\bar1}\wedge\beta^1+\bb^{\bar2}\wedge\beta^2\big) \ , 
%\label{adbetam13}
\displaybreak[0]\\[4pt]
\bb_1{}^\dag\wedge\bb_1
=&~\frac32\,\begin{pmatrix}
\beta^2\wedge\bb^{\bar2}&-\beta^2\wedge\bb^{\bar1}\\-\beta^1\wedge\bb^{\bar2}&
\beta^1\wedge\bb^{\bar1}\end{pmatrix} \ , %
%\label{adbetam13dag}
\displaybreak[0]\\[4pt]
\bb_2\wedge\bb_2{}^\dag
=&~\begin{pmatrix}
\bb^{\bar1}\wedge\beta^1&\sqrt{\frac12}~\bb^{\bar1}\wedge\beta^2&0\\
\sqrt{\frac12}~\bb^{\bar2}\wedge\beta^1&\frac12\,\big(
\bb^{\bar1}\wedge\beta^1+\bb^{\bar2}\wedge\beta^2\big)&\sqrt{\frac12}~
\bb^{\bar1}\wedge\beta^2\\0&\sqrt{\frac12}~\bb^{\bar2}\wedge\beta^1&
\bb^{\bar2}\wedge\beta^2\end{pmatrix} \ , \label{adbetap13}\\[4pt]
 \displaybreak[0]
\bb_2{}^\dag\wedge\bb_2
=&~\begin{pmatrix}
\beta^1\wedge\bb^{\bar1}+\frac12\,\beta^2\wedge\bb^{\bar2}&\frac12\,
\beta^2\wedge\bb^{\bar1}\\\frac12\,\beta^1\wedge\bb^{\bar2}&
\frac12\,\beta^1\wedge\bb^{\bar1}+\beta^2\wedge\bb^{\bar2}\end{pmatrix} \ ,
%\label{adbetap13dag}
\displaybreak[0]\\[4pt]
\bb_3\wedge\bb_1
=&~\mbox{$\frac32$}\,\bb^{\bar1}\wedge\bb^{\bar2}~\idm_2 \ , 
%\label{adbeta0013}
\displaybreak[0]\\[4pt]
\bb_1\wedge\bb_2{}^\dag
=&~\sqrt{\mbox{$\frac32$}}~
\Big(\bb^{\bar2}\wedge\beta^1 \ , \ -\sqrt{\mbox{$\frac12$}}~\big(
\bb^{\bar1}\wedge\beta^1+\bb^{\bar2}\wedge\beta^2\big) \ , \ 
-\bb^{\bar1}\wedge\beta^2\Big) \ .
%\label{adbetapm13}
\end{align*}
\paragraph{Yang-Mills equations.}
We are now prepared to derive the equation of motion for the Higgs fields from the Yang-Mills equations for the gauge potential (\ref{ansatz c11 cp2 gauge potential}) and field strength (\ref{ansatz c11 cp2 gauge field strength}).
For the calculations, we are going to use the Levi-Civita connection 1-form from (\ref{connection 1-form cp2}) as well as the non-vanishing structure constants given in (\ref{structure constants cp2}).

The form of the Yang-Mills equations does not change and is given by (\ref{YM cp2 c20 1}) and (\ref{YM cp2 c20 2})
% \begin{eqnarray}
%  \CD_a \CF^{a0} &=& 0\,, \label{YM cp2 c11 1}\\
%  \CD_0 \CF^{0b} + \CD_a \CF^{ab} &=& 0 \label{YM cp2 c11 2}
% \end{eqnarray}
with the same notation as before. Again, we split the gauge potential into its block-diagonal and off-diagonal part. If we insert the gauge potential (\ref{ansatz c11 cp2 gauge potential}) into (\ref{YM cp2 c20 1}), we recognize again that the left hand side vanishes and does not restrict our scalar fields. The other set of equations (\ref{YM cp2 c20 2}) with a free coset superscript again decomposes to equation (\ref{ansatz cp1 c20 yang mills equation2}).
% \begin{equation}\label{ansatz cp1 c11 yang mills equation}
%  0=\frac{\dr}{\dr\tau} \CF^{0b}-e^i{}_a\left(f_{ci}{}^a~\CF^{cb}+ f_{ci}{}^b~\CF^{ac} \right)+\left[\CA_a^{\text{diag}},\CF^{ab}\right] + \left[\CA_a^{\text{off}},\CF^{ab}\right]\,.
% \end{equation}
By inserting (\ref{ansatz c11 cp2 gauge potential}), (\ref{ansatz c11 cp2 gauge field strength}) as well as (\ref{structure constants cp2}) and (\ref{e^i_b cp2}) into (\ref{ansatz cp1 c20 yang mills equation2}) and after a fair amount of calculations, we find that equation (\ref{ansatz cp1 c20 yang mills equations vanishing piece})
% \begin{equation}
%  0 = e^i{}_a\left(f_{ci}{}^a~\CF^{cb}+ f_{ci}{}^b~\CF^{ac} \right) - \left[\CA_a^{\text{diag}},\CF^{ab}\right]\,,
% \end{equation}
which was a trivial condition for the case of two scalar fields, is not trivial here, but restricts our fields by the equation
\begin{equation}\label{ansatz cp2 c11 diag equation}
 \phi_1~\phi_2=\phi_3~\phi_4\,.
\end{equation}
Algebraically equation (\ref{ansatz cp2 c11 diag equation}) represents the relation of the quiver expressing commutativity of the quiver diagram. The remaining part of (\ref{ansatz cp1 c20 yang mills equation2}) yields (\ref{ansatz cp1 c20 yang mills equations non vanishing piece})
% \begin{equation}
%   0=\frac{\dr}{\dr\tau} \CF^{0b} + \left[\CA_a^{\text{off}},\CF^{ab}\right]\,,
% \end{equation}
which for every index $b$ becomes a matrix equation containing six differential equations for the four scalar fields. For $\be=1,~\bar1$, four out of six equations turn out to be independent. With free coset index $2$ or $\bar2$, we get six independent equations. If we make use of the algebraic constraint, these six reduce to a system of four independent equations that coincide with the ones we found for $b=1$ and $b=\bar1$ and read:
\begin{subequations}
\begin{eqnarray}
\label{ansatz cp2 c11 eom1}
  \frac{\dr^2}{\dr\tau^2}~\phi_1
     &=& 3~\phi_1\left(\frac{3}{2}~\phi_1^2 - 1 - \frac12~\phi_2^2\right)\,, \\
\label{ansatz cp2 c11 eom2}
\frac{\dr^2}{\dr\tau^2}~\phi_2
     &=& 3~\phi_2\left(\frac56~\phi_2^2 - 1-\frac12~\phi_1^2 + \frac23 ~\phi_4^2 \right)\,,\\
\label{ansatz cp2 c11 eom3}
 \frac{\dr^2}{\dr\tau^2}~\phi_3
     &=& 3~\phi_3\left(\frac32~\phi_3^2 - 1 -\frac12 \phi_4^2\right)\,,\\
\label{ansatz cp2 c11 eom4}
\frac{\dr^2}{\dr\tau^2}~\phi_4
     &=& 3~\phi_4\left(\frac56 \phi_4^2 - 1 -\frac12 \phi_3^2 +\frac23 ~\phi_2^2\right)\,.
\end{eqnarray}
\label{ansatz cp2 c11 eom}
\end{subequations}
It is not easy to solve these equations, but we can get simplifications under certain conditions. In order to do that we can either put some fields to zero, or simply identify two fields with each other. Due to the algebraic condition (\ref{ansatz cp2 c11 diag equation}) is it not possible that only one field equals zero. We have six cases in which two fields are equal and then we can employ the algebraic condition (\ref{ansatz cp2 c11 diag equation}) to either set them zero or to equate the remaining two fields. So there are twelve possibilities and we find that six of them actually force all fields to be the same, namely $\phi_1=\phi_2=\phi_3=\phi_4$. In this case the differential equations (\ref{ansatz cp2 c11 eom1})-(\ref{ansatz cp2 c11 eom4}) simplify to the equation (\ref{ansatz cp2 c20 eom 3}).
%But this is clear if one takes a look at the generic ansatz in \cite{preprint5}
The remaining four independent possibilities simplify (\ref{ansatz cp2 c11 eom1})-(\ref{ansatz cp2 c11 eom4}) as follows:
\begin{enumerate}
 \item $\phi_1 = \phi_3 = 0,~\phi_2\ne 0,~\phi_4\ne 0$:
\begin{eqnarray}\label{ansatz cp2 c11 eom5}
 \frac{\dr^2}{\dr \tau^2}\phi_2=3\,\phi_2 \left( \frac56\, \phi_2^2-1+\frac23\,\phi_4^2\right)\,, \qquad
 \frac{\dr^2}{\dr \tau^2}\phi_4=3\,\phi_4 \left( \frac56\, \phi_4^2-1+\frac23\,\phi_2^2 \right)\,.
\end{eqnarray}
 \item $\phi_1 = \phi_3 \ne 0$,~$\phi_2 = \phi_4 \ne 0$:
\begin{eqnarray}\label{ansatz cp2 c11 eom6}
 \frac{\dr^2}{\dr \tau^2}\phi_1=3\,\phi_1 \left( \frac32\, \phi_1^2-1-\frac12\,\phi_2^2\right)\,, \qquad
 \frac{\dr^2}{\dr \tau^2}\phi_2=3\,\phi_2 \left( \frac32\, \phi_2^2-1-\frac12\,\phi_1^2 \right).
\end{eqnarray}
% 
% 
% for instance the the following set of solutions
% \begin{eqnarray}
%  \phi_1(\tau) = \sqrt{\frac23}\tanh{\left(\sqrt{\frac32}\tau\right)},~\phi_2(\tau) = 0 ~~~\text{or}~~~
%  \phi_2(\tau) = \sqrt{\frac23}\tanh{\left(\sqrt{\frac32}\tau\right)},~\phi_1(\tau) = 0\,.
% \end{eqnarray}
 \item $\phi_1 = \phi_4 = 0$,~$\phi_2 \ne 0,~\phi_3 \ne 0$:
\begin{eqnarray}\label{ansatz cp2 c11 eom7}
 \frac{\dr^2}{\dr \tau^2}\phi_2=3\,\phi_2 \left( \frac56\, \phi_2^2-1\right)\,, \qquad
 \frac{\dr^2}{\dr \tau^2}\phi_3=3\,\phi_3 \left( \frac32\, \phi_3^2-1\right)\,.
\end{eqnarray}

\item $\phi_2 = \phi_3 = 0$,~$\phi_1 \ne 0,~\phi_4\ne 0$:
\begin{eqnarray}\label{ansatz cp2 c11 eom8}
 \frac{\dr^2}{\dr \tau^2}\phi_1=3\,\phi_1 \left( \frac32\, \phi_1^2-1\right)\,, \qquad
 \frac{\dr^2}{\dr \tau^2}\phi_4=3\,\phi_4 \left( \frac56\, \phi_4^2-1\right)\,.
\end{eqnarray}

% \item ($\phi_2 = \phi_4 = 0$)
% \begin{eqnarray}\label{ansatz cp2 c11 eom9}
%  \frac{\dr^2}{\dr \tau^2}\phi_1=3\,\phi_1 \left( \frac32\, \phi_1^2-1\right)\,, \qquad
%  \frac{\dr^2}{\dr \tau^2}\phi_3=3\,\phi_3 \left( \frac32\, \phi_3^2-1\right)\,.
% \end{eqnarray}

\end{enumerate}
Here, the decoupled equations (\ref{ansatz cp2 c11 eom7}) and (\ref{ansatz cp2 c11 eom8}) are similar to those we found before in (\ref{ansatz cp2 c20 eom 4}) and (\ref{ansatz cp2 c20 eom 5}) and, for instance, can be solved by
\begin{eqnarray}
 \phi_1(\tau) = 0,\quad \phi_2(\tau) = \sqrt{\frac65}\tanh{\left(\sqrt{\frac32}~\tau \right)},\quad  \phi_3(\tau) = \sqrt{\frac23}\tanh{\left(\sqrt{\frac32}~\tau\right)},\quad \phi_4(\tau)=0\,,\\
 \phi_1(\tau) = \sqrt{\frac23} \tanh{\left(\sqrt{\frac32}~\tau\right)},\quad \phi_2(\tau)=0,\quad \phi_3(\tau) = 0, \quad\phi_4(\tau) = \sqrt{\frac65}\tanh{\left(\sqrt{\frac32}~\tau\right)}\,.
 \end{eqnarray}

\begin{figure}[ht]
\subfigure[Solutions of (\ref{ansatz cp2 c11 eom}) with the following initial values: \newline
$\phi_1(0)=0.3,~\dot\phi_1(0)=0,~\phi_2(0)=0.4,~\dot\phi_2(0)=0,~
\newline \phi_3(0)=0,~~~\dot\phi_3(0)=1,~\phi_4(0)=0,~~~\dot\phi_4(0)=1.2$
]{
  \includegraphics[width=7.9cm]{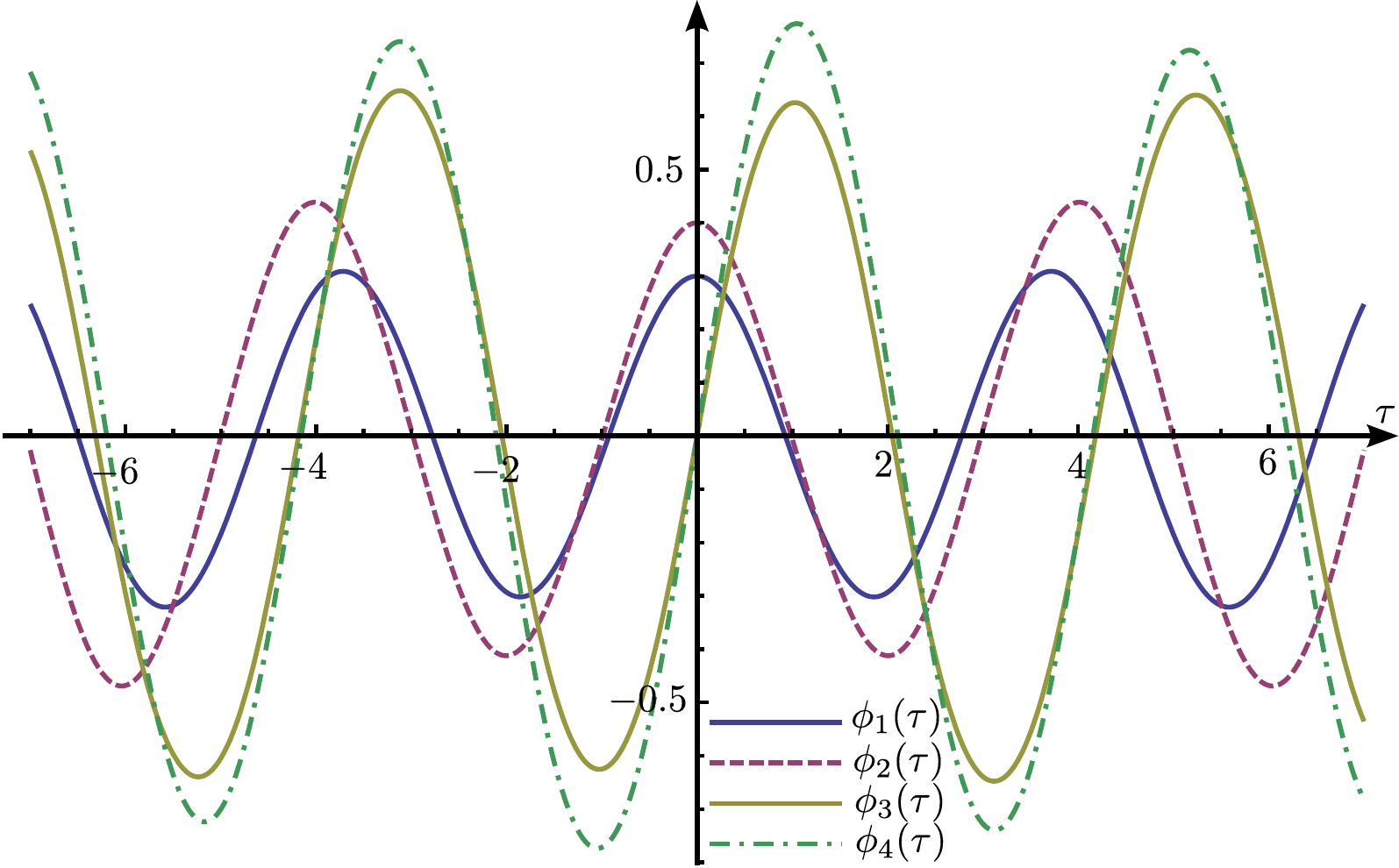}
}
\subfigure[Solutions of (\ref{ansatz cp2 c11 eom}) with the following initial values:\newline
 $\phi_1(0)=0,~\dot\phi_1(0)=1.1,~\phi_2(0)=0.4,~\dot\phi_2(0)=0.3,~
\newline\phi_3(0)=0,~\dot\phi_3(0)=1,~~~\phi_4(0)=1,~~~\dot\phi_4(0)=0$
]{
 \includegraphics[width=7.9cm]{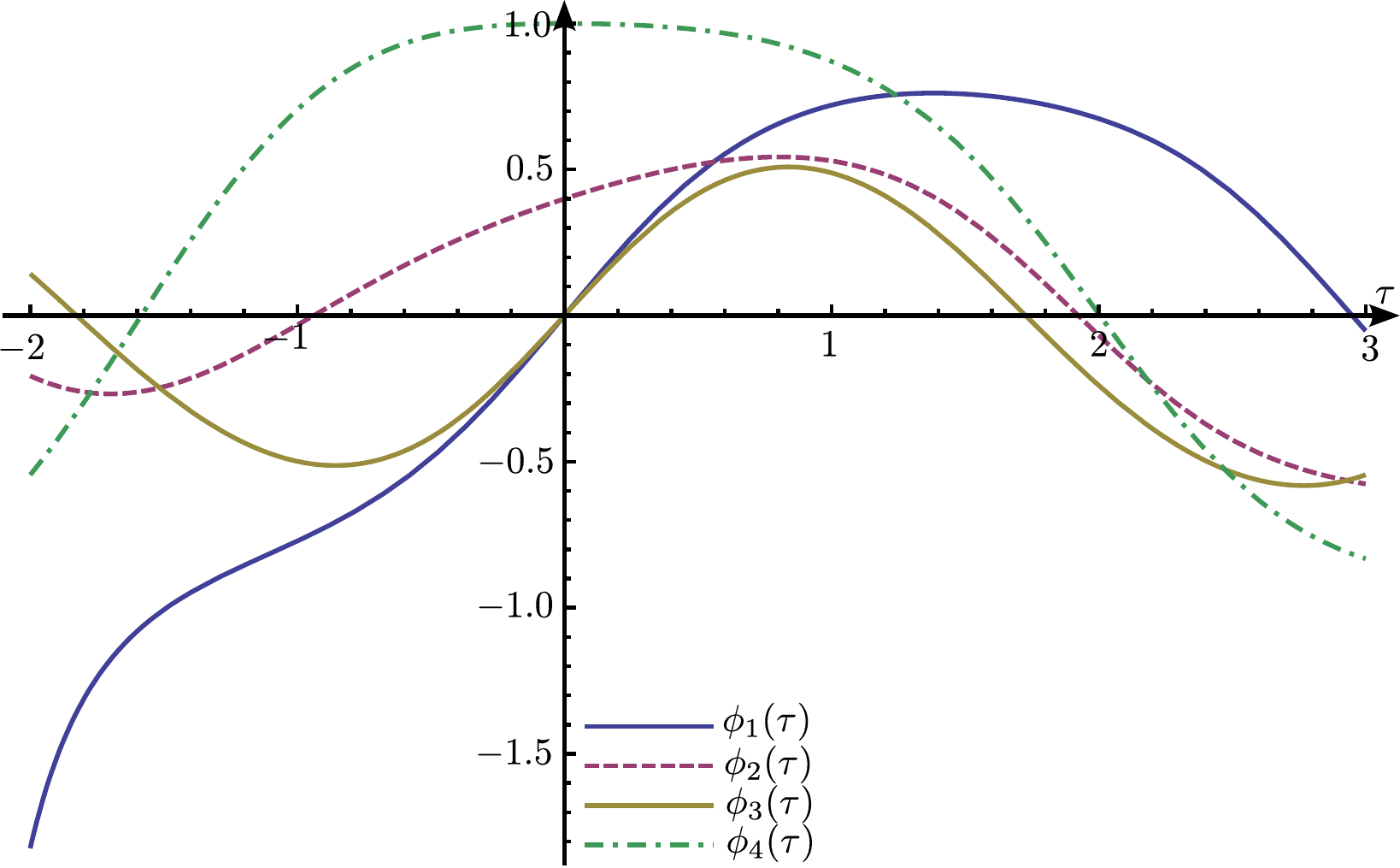}
}
  \caption{\small Two different sets of truly coupled numerical solutions $\phi_1(\tau),~\phi_2(\tau),~\phi_3(\tau),~\phi_4(\tau)$ to the system of differential equations (\ref{ansatz cp2 c11 eom}). Particularly $\phi_i\neq\phi_j,~\phi_i\neq0,~\forall~i,j\in\{1,2,3,4\} $.}
  \label{fig:RxCP2-C11}
\end{figure}

Truly coupled solutions to (\ref{ansatz cp2 c11 eom}) are again quite hard to find, but we could still manage to find numerical solutions to the system with different initial values, stated in figure \ref{fig:RxCP2-C11}.

\section[Yang-Mills theory on R x Q3 in C(1,0) quiver representation]{Yang-Mills theory on $\mathbb R \times Q_3$ in $\underline{C}^{1,0}$ quiver representation}
%\section{YANG-MILLS THEORY ON $\mathbb R \times Q_3$ IN $\underline{C}^{1,0}$ QUIVER REPRESENTATION}
\addcontentsline{toc}{section}{Yang-Mills on R x Q3 in C(1,0) quiver representation}
Now we turn our attention to a different coset space, namely $G/H=Q_3$, which is the quotient
\begin{equation*}
 Q_3 := \frac{SU(3)}{U(1)\times U(1)}\,,
\end{equation*}
and which is a homogeneous but not symmetric space in contrast to the case of $\BC P^2$.
\paragraph{Invariant 1-forms on \boldmath{$Q_3$}.}
As in section \ref{Invariant 1-forms on CP2} we first want to write down the invariant 1-forms on $Q_3$ and the $SU(3)$-equivariant gauge potential for the fundamental representation of $SU(3)$. These are taken from \cite{preprint5}, and the explicit derivation can be looked up there.

For the space $Q_3$ with three complex dimensions we have six linearly independent invariant 1-forms. The explicit form of these 1-forms is given in (3.39) from \cite{preprint5}. They are denoted by
\begin{equation*}
 \left\{e^1,~e^2,~e^3,e^{\bar1},~e^{\bar2},~e^{\bar3} \right\} =: \left\{\ga^1,~\ga^2,~\ga^3,~\bg^{\bar1},~\bg^{\bar2},~\bg^{\bar3}\right\}\,.
\end{equation*}
Since the subgroup $H=U(1)\times U(1)$ is a different than before, we get a different decomposition of the irreducible representation of $SU(3)$. This means that if we choose the fundamental representation of $SU(3)$ then as the simplest case we get the following decomposition:
\begin{equation}\label{ansatz q1 decomposition h}
\underline{C}^{1,0}\big|_{U(1)\times U(1)}=\underline{(1,1)_1}~\oplus~
\underline{(-1,1)_1}~\oplus~\underline{(0,-2)_0}\,.
\end{equation}
Here in the $\underline{(q,m)_n}$ the $n$ are the same as explained for the symmetric case (\ref{decomposition1CP1}), representing the original $SU(3)$ isospin. The pairs $(q,m)$ denote magnetic charges of two $U(1)$ subgroups of $SU(3)$ which can be read off from the eigenvalues of the generators $E_7,~E_8$ in (\ref{generators H Q3}) below. Since $n$ equals twice the isospin, we find that $q=-n,-n+2,...,n-2,n$ equals just two times the third component of the isospin.
As one can see here, we already have a decomposition into three irreps of $H$. The corresponding quiver diagram shows that there will appear three independent scalar fields in the $SU(3)$-equivariant gauge potential on the corresponding associated quiver bundle:
\begin{equation*}
\xymatrix{ \BR\otimes \CV_{(-1,1)}^{Q_3} \ar[rr]^{\phi_3\otimes\gamma_3}&  & \BR\otimes \CV_{(1,1)}^{Q_3} \\
	   &\ar[lu]^{\phi_2\otimes\gamma_2}\BR\otimes \CV_{(0,-2)}^{Q_3}\ar[ru]_{\phi_1\otimes\gamma_1}&} 
\end{equation*}
Due to the fact that each term in (\ref{ansatz q1 decomposition h}) corresponds to a 1-dimensional representation of $H$, the structure group for the associated vector bundle is $U(3)$.

The generators corresponding to $Q_3$ in this representation are then given by
\begin{equation*}
 \begin{aligned}
 &E_1 = e_{31}\,,\qquad & E_2 = e_{32}\,,\qquad & E_3 =e_{21}\,,\\
 &E_{\bar1} = e_{13}\,,\qquad & E_{\bar2} = e_{23}\,,\qquad & E_{\bar3}= e_{12}\,, \\
 \end{aligned}
\end{equation*}
along with the generators of $H$,
\begin{equation}\label{generators H Q3}
 E_7 = e_{11}-e_{22},\qquad E_8=e_{11}+e_{22}-2e_{33}\,.
\end{equation}
Hence, we have the following structure constants:
 \begin{align*}
  &f_{ab}{}^c:& & f_{\bar3\bar2}{}^{\bar1}=+1 & \quad  & f_{\bar31}{}^{2}=-1 & \quad  & f_{\bar21}{}^{3}=+1 \notag\\
  & & 	        &f_{32}{}^{1}=-1 & \quad  & f_{3\bar1}{}^{\bar2}=+1 & \quad  & f_{2\bar1}{}^{\bar3}=-1 \displaybreak[0]\notag\\
&f_{ai}{}^c:& & f_{\bar37}{}^{\bar3}=-2 & \quad  & f_{\bar27}{}^{\bar2}=+1 & \quad  & f_{\bar17}{}^{\bar1}=-1\notag\\
  & & 	        &f_{37}{}^{3}=+2 & \quad  & f_{27}{}^{2}=-1 & \quad  & f_{17}{}^{1}=+1 \notag\\
&	    & & f_{\bar38}{}^{\bar3}=0 & \quad  & f_{\bar28}{}^{\bar2}=-3 & \quad  & f_{\bar18}{}^{\bar1}=-3\notag\\
  & & 	        &f_{38}{}^{3}=0 & \quad  & f_{28}{}^{2}=+3 & \quad  & f_{18}{}^{1}=+3\,.
\end{align*}
\paragraph{Gauge potential and field strength.}
Next we want to write down the flat connection on the trivial $\BC^3$-bundle over $Q_3$ what we also first did for the $\BC P^2$ case in (\ref{ansatz cp2 c10 flat connection}). The flat connection on the trivial bundle over $Q_3$ is given in the invariant basis as
\begin{equation}\label{flat connection c10 Q3}
A_0 =~\begin{pmatrix}
a_1&\bg^{\bar3}&\bg^{\bar1}\\[4pt]
-\ga^3&-a_1-a_2&\bg^{\bar2}\\[4pt]
-\ga^1&-\ga^2& a_2\end{pmatrix} \,.
\end{equation}
Here, $a_1$ and $a_2$ are $u(1)$-valued connection 1-forms given in equation (3.38) from \cite{preprint5}.
The remaining invariant 1-forms $e^7,~e^8$ on $G/H$ correspond to the Lie algebra Lie$(H)$ and can be written in terms of $U(1)\times U(1)$ gauge potentials $a_1$ and $a_2$.
%of $G$ decompose as 1-forms on $G/H$ and can be written in terms of the $a_1,~a_2$ where we denote the 1-forms on $H$ by $e^7,~e^8$. 
They have the following components:
\begin{equation*}
 e^7{}_b = \left(a_1+\frac12 a_2\right)_b \qquad e^8{}_b = -\frac12 (a_2)_b\,.
\end{equation*}

The flat connection (\ref{flat connection c10 Q3}) satisfies the Maurer-Cartan equations 
\begin{equation}
 \dr A_0 + A_0\wedge A_0 = 0
\end{equation}
which yields the following equations for the invariant one-forms and $u(1)$-valued connection 1-forms:
\begin{subequations}
\begin{eqnarray}
\label{ansatz Q1 c10 maurer cartan 1}
\dr a_1 -\bg^1\wedge\gamma^1 -\bg^3\wedge\gamma^3 &=& 0\,, \\
\dr a_2 +\bg^1\wedge\gamma^1+\bg^2\wedge\gamma^2 &=& 0\,, \\
\dr\gamma^1-(a_1-a_2)\wedge\gamma^1-\gamma^2\wedge\gamma^3 &=& 0 \,,\\
\dr\gamma^2+(a_1+2a_2)\wedge\gamma^2+\gamma^1\wedge\bg^3 &=& 0 \,,\\
\label{ansatz Q1 c10 maurer cartan 5}
\dr \gamma^3-(2a_1+a_2)\wedge\gamma^3-\gamma^1\wedge\bg^2 &=& 0\,.
\end{eqnarray}
\end{subequations}

The extension to the non-flat connection on the corresponding extended bundle, taken from (3.50) in \cite{preprint5}, reads
\begin{equation}\label{ansatz Q1 c10 gauge potential}
 \CA=\begin{pmatrix}
  a_1 & \phi_3~\bg^{\bar3} & \phi_1~\bg^{\bar1}\\
  -\phi_3~\ga^3 & -a_1-a_2 & -\phi_2~\bg^{\bar2}\\
  -\phi_1~\ga^1 & -\phi_2~\ga^2 & a_2
 \end{pmatrix}\,.
\end{equation}
The corresponding field strength $\CF=\dr\CA+\CA\wedge\CA$ is then easily calculated using the equations (\ref{ansatz Q1 c10 maurer cartan 1})-(\ref{ansatz Q1 c10 maurer cartan 5}):
%
%\begin{small}
\begin{equation}\label{ansatz Q1 c10 gauge field}
  \CF=
\left(\begin{aligned}
    & \begin{matrix} (1-\phi_1^2)\bg^{\bar1}\wedge\ga^1\\ +~(1-\phi_3^2)\bg^{\bar3}\wedge\ga^3 \end{matrix}&
    & \begin{matrix} \dr\phi_3 \wedge \bg^{\bar3}\\ +~(\phi_3-\phi_1\phi_2)~\bg^{\bar1}\wedge\ga^2 \end{matrix}&
    & \begin{matrix} \dr\phi_1 \wedge \bg^{\bar1}\\ +~(\phi_1-\phi_3\phi_2)~\bg^{\bar2}\wedge\bg^{\bar3} \end{matrix}& \\[0.3cm]
    & \begin{matrix} -\dr\phi_3 \wedge \ga^3\\ -~(\phi_3-\phi_1\phi_2)~\ga^{1}\wedge\bg^{\bar2} \end{matrix}&
    & \begin{matrix} -(1-\phi_3^2)~\bg^{\bar3}\wedge\ga^3\\ +~(1-\phi_2^2)~\bg^{\bar2}\wedge\ga^2 \end{matrix}&
    & \begin{matrix} \dr\phi_2\wedge\bg^{\bar2}\\ +~(\phi_2-\phi_3\phi_1)~\ga^3\wedge\bg^{\bar1} \end{matrix}&\\[0.3cm]
    & \begin{matrix} -\dr\phi_1 \wedge \ga^1\\ -~(\phi_1-\phi_3\phi_2)~\ga^2\wedge\ga^3 \end{matrix}&
    & \begin{matrix} -\dr\phi_2\wedge\ga^2\\ -~(\phi_2-\phi_3\phi_1)~\bg^{\bar3}\wedge\ga^1 \end{matrix}&
    & \begin{matrix} -(1-\phi_1^2)~\bg^{\bar1}\wedge\ga^1\\ -~(1-\phi_2^2)~\bg^{\bar2}\wedge\ga^2 \end{matrix}&
 \end{aligned}\right)\,.
\end{equation}
%\end{small}
%
\paragraph{Yang-Mills equations.}
The Yang-Mills equations on $Q_3$ look a little different in this case, since we have another set of non-vanishing structure constants, namely those with coset indices. We can therefore endow $Q_3$ with a non-vanishing torsion tensor with non-holonomic components
\begin{equation}
 T^b_{ac}=\kappa~f_{ac}{}^b\,.
\end{equation}
Such a torsion tensor was introduced in a similar way in \cite{yangmillsflow}. We end up with the following Yang-Mills equations:
\begin{eqnarray}\label{ansatz Q1 yang-mills eqn}
\text{YM}^b &:=&\frac{\dr}{\dr\tau} \CF^{0b}+\frac{(1+\kappa)}{2}\left(f_{ac}{}^b \CF^{ac}+f_{ac}{}^a \CF^{cb}\right)\nonumber\\
& & -e^i{}_a\left(f_{ci}{}^a~\CF^{cb}+ f_{ci}{}^b~\CF^{ac} \right)+\left[\CA_a,\CF^{ab}\right]= 0\,.
\end{eqnarray}

If we insert (\ref{ansatz Q1 c10 gauge potential}) and (\ref{ansatz Q1 c10 gauge field}) into equation (\ref{ansatz Q1 yang-mills eqn}), we find an independent differential equation for the scalar fields $\phi_1,~\phi_2,~\phi_3$ for each superscript $b=1,~2,~3,~\bar1,~\bar2,~\bar3$. These six equations actually differ via three algebraic conditions on the fields which come from the term containing the coset structure constants in (\ref{ansatz Q1 yang-mills eqn}). We can separate the algebraic conditions by adding and subtracting those equations that have conjugated indices, schematically
\begin{equation*}
  \text{YM}^1\pm\text{YM}^{\bar1}, \qquad \text{YM}^2\pm\text{YM}^{\bar2}, \qquad \text{YM}^3\pm\text{YM}^{\bar3}\,.
\end{equation*}
By doing that, we arrive at three independent differential equations
\begin{subequations}
\begin{eqnarray}
\label{ansatz Q1 c10 eom1}
 \frac{\dr^2}{\dr\tau^2} \phi_1 &=& 2\phi_1~\left(\phi_1^2-1+\frac12\left(\phi_2^2+\phi_3^2\right)\right)-2\phi_2~\phi_3\,,
\\
\label{ansatz Q1 c10 eom2}
 \frac{\dr^2}{\dr\tau^2} \phi_2 &=& 2\phi_2~\left(\phi_2^2-1+\frac12\left(\phi_1^2+\phi_3^2\right)\right)-2\phi_1~\phi_3\,,
\\
\label{ansatz Q1 c10 eom3}
 \frac{\dr^2}{\dr\tau^2} \phi_3 &=& 2\phi_3~\left(\phi_3^2-1+\frac12\left(\phi_1^2+\phi_2^2\right)\right)-2\phi_1~\phi_2\,,
\end{eqnarray}
\label{ansatz Q1 c10 eom}
\end{subequations}
along with the algebraic constraints
\begin{subequations}
\label{ansatz Q1 c10 algebraic constraints}
\begin{eqnarray}
 (\kappa+1)(\phi_1-\phi_2~\phi_3) &=& 0\,,\\
 (\kappa+1)(\phi_2-\phi_1~\phi_3) &=& 0\,,\\
 (\kappa+1)(\phi_3-\phi_1~\phi_2) &=& 0\,.
\end{eqnarray}
\end{subequations}

From (\ref{ansatz Q1 c10 algebraic constraints}) it follows that for $\kappa\neq-1$ the Higgs fields are constrained by the relations of the pertinent quiver which restrict us to locally constant fields with values $1,~0,~-1$. For $$(\phi_1,\phi_2,\phi_3)\in\left\{(1,1,1),(-1,1,1),(1,-1,1),(1,1,-1),(-1,-1,1),(-1,1,-1),(1,-1,-1)\right\}$$ we find that the gauge connection (\ref{ansatz Q1 c10 gauge potential}) is flat but for $(\phi_1,\phi_2,\phi_3)=(0,0,0)$ it is not flat and solves the Yang-Mills equations on $Q_3$. For $\kappa=-1$, the constraints (\ref{ansatz Q1 c10 algebraic constraints}) are resolved for any $(\phi_i)$, i.\ e.\ for a specific torsion in the torsionful Yang-Mills equations the terms responsible for quiver relations cancel one another. %In order to satisfy the algebraic constraints, we are forced to choose the value $\kappa=-1$ for the torsion. Otherwise these conditions would restrict us to constant fields with values $1,~0$ or $-1$.
So, if we choose the specific value $\kappa=-1$ for the torsion, (\ref{ansatz Q1 c10 eom1})-(\ref{ansatz Q1 c10 eom3}) can in principle be solved. It is not very easy in general but some solutions to these equations can be obtained by putting two out of three fields to zero which yields
%it is not very easy. Nevertheless, possible solutions for these equations are obtained at least by putting two out of three fields to zero which yields
\begin{eqnarray}
 \frac{\dr^2}{\dr\tau^2} \phi_i &=& 2\phi_i~\left(\phi_i^2-1\right),~\text{if}\quad \phi_j = 0~\forall~j \neq i\,,\quad i,j\in\left\{1,2,3\right\}
 %\frac{\dr^2}{\dr\tau^2} \phi_2 &=& 2\phi_2~\left(\phi_2^2-1\right),\qquad \text{for}\quad \phi_1 = \phi_3 = 0\,,\\
 %\frac{\dr^2}{\dr\tau^2} \phi_3 &=& 2\phi_3~\left(\phi_3^2-1\right),\qquad \text{for}\quad \phi_1 = \phi_2 = 0\,.
\end{eqnarray}
and is solved by
\begin{equation}
 \phi_i=\tanh(\tau),~\phi_j=0 \quad\forall~j \neq i\,,\quad i,j\in\left\{1,2,3\right\}\,.
\end{equation}
\begin{figure}[ht]
\subfigure[Solutions  of (\ref{ansatz Q1 c10 eom}) with the following initial values:\newline $\phi_1(0)=0,~\dot\phi_1(0)=1,~\phi_2(0)=0,~\dot\phi_2(0)=0.5,~
\newline \phi_3(0)=1,~\dot\phi_3(0)=0$
]{
  \includegraphics[width=7.9cm]{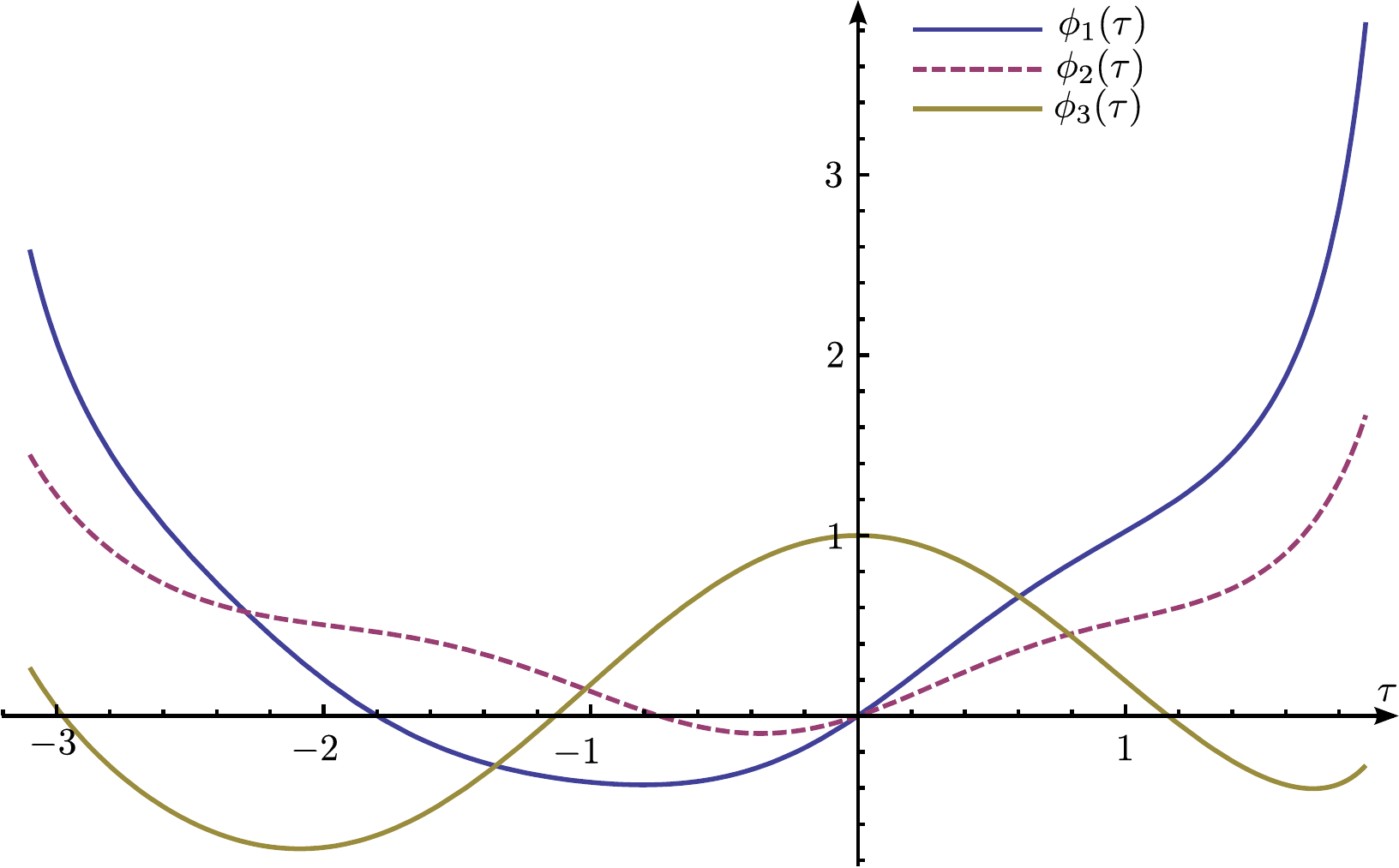}
}
\subfigure[Solutions of (\ref{ansatz Q1 c10 eom}) with the following initial values:\newline $\phi_1(0)=1.05,~\dot\phi_1(0)=0,~\phi_2(0)=1.1,~\dot\phi_2(0)=0,~
\newline \phi_3(0)=1.2,~~\dot\phi_3(0)=0$
]{
 \includegraphics[width=7.9cm]{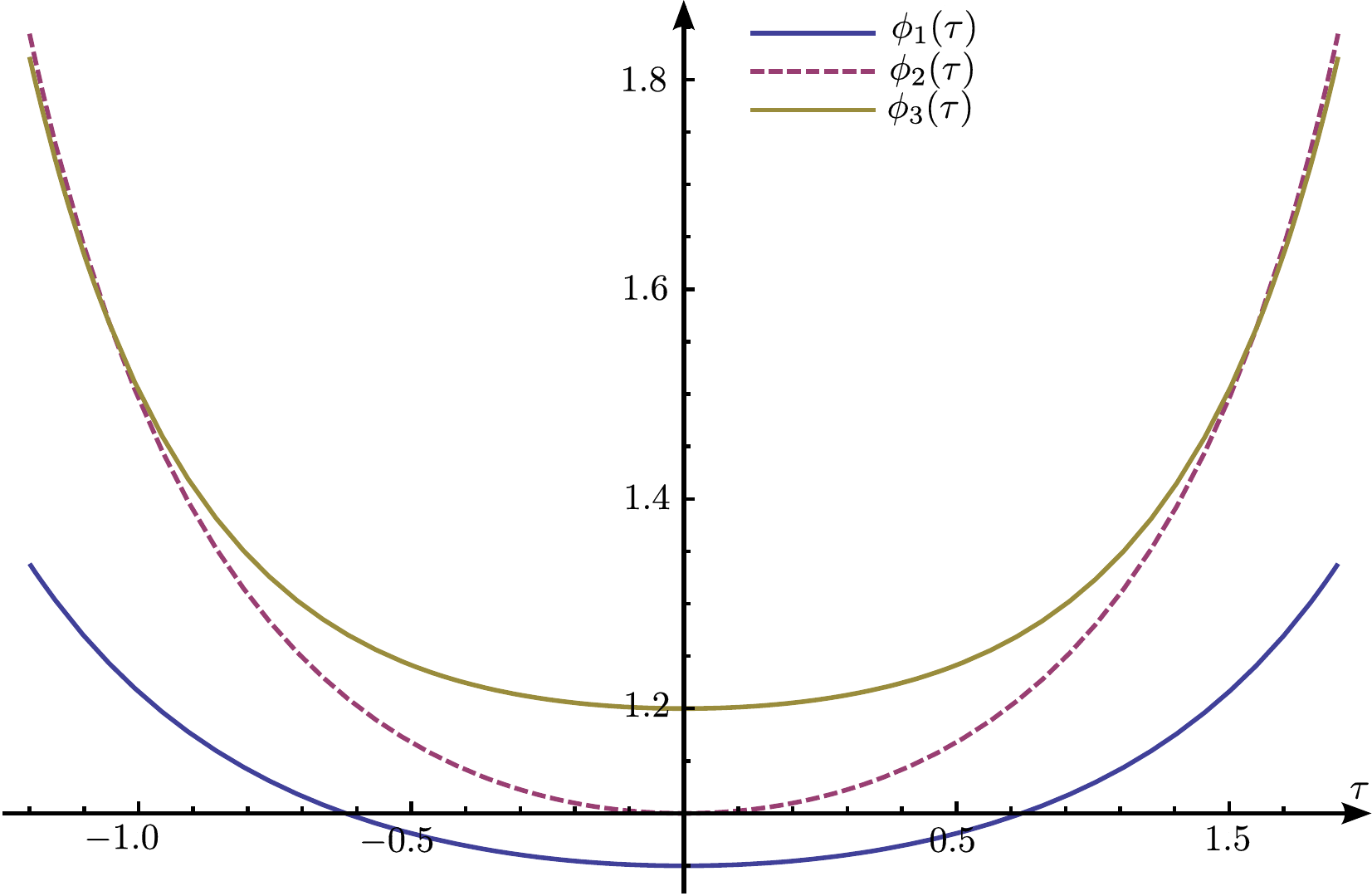}
}
  \caption{\small Two different sets of truly coupled numerical solutions $\phi_1(\tau),~\phi_2(\tau),~\phi_3(\tau)$ to the system of differential equations (\ref{ansatz Q1 c10 eom}). Particularly $\phi_i\neq\phi_j,~\phi_i\neq0,~\forall~i,j\in\{1,2,3\} $.}
  \label{fig:RxQ3-C10}
\end{figure}
Also for this system of differential equations (\ref{ansatz Q1 c10 eom}), truly coupled solutions could be found numerically for certain initial  values and are stated in figure \ref{fig:RxQ3-C10}.
%
%But as in the case of $\BC P^2$ in the last section, we do not want to go further in the analysis here but simply finish all our considerations with this result.
The next thing one could do would be to choose higher representations such as $\underline{C}^{2,0}$ or $\underline{C}^{1,1}$ for $SU(3)$ and get further decompositions by restricting to the subgroup $H=U(1)\times U(1)$. Therefore more scalar fields would arise for the corresponding $SU(3)$-equivariant ans\"atze. But we want to stop the analysis of quiver bundles over $Q_3$ at this point and turn to a different space.
\section[Yang-Mills theory on R x CP1 x CP2]{Yang-Mills theory on $\BR \times \BC P^1\times \BC P^2$}
%\section{YANG-MILLS THEORY ON $\BR \times \BC P^1\times \BC P^2$}
\addcontentsline{toc}{section}{Yang-Mills on R x CP1 x CP2}
\label{yang mills on RxCP1xCP2}
In the following we want to consider a more general situation, where the base space is $\BR \times \BC P^1\times \BC P^2$ and a vector bundle over it has the structure group $U(3m+3)$. Therefore, in contrast to all other examples, we get a more general ansatz for the corresponding gauge potential on the associated vector bundle which contains $2m+1$ scalar fields. Here we do not want to fix the specific number of Higgs fields, but derive the equation of motion for this ansatz for arbitrary $m$.
\paragraph{The ansatz for a gauge potential.}
We are making an ansatz for a $u(3m+3)$-valued gauge potential (in the temporal gauge $\CA_\tau=0$), which is a modified combination of the ans\"atze, taken in \cite{preprint5} and \cite{preprint1,preprint2}, such that
\begin{eqnarray}
\label{ansatz cp1cp2 gauge potential}
 \CA &=& A_m\otimes \idm_3+\idm_{m+1}\otimes \left(\begin{matrix} B&0\\ 0&-2a\end{matrix}\right)
 + \Psi_m \otimes \left(\begin{matrix} 0_2&\bar\beta \\ -\beta^\top &0\end{matrix}\right)\,,
\end{eqnarray}
where
\begin{eqnarray*}
 B := B_{(1)}+a\cdot\idm_2\,.
\end{eqnarray*}
As before, $B_{(1)}$ denotes the $su(2)$-valued one-instanton field on $\BC P^2$ and the $\be,~\bb$ are row vectors of the invariant basis of 1-forms of $\BC P^2$, given in (\ref{left invariant cp2 1}),(\ref{left invariant cp2 2}),
\begin{equation*}
 \Psi_m:=\text{diag}(\psi_1,\psi_2,...,\psi_{m+1})\,,
\end{equation*}
where all $\psi_i$ are considered to be real scalar fields on $\BR$. Furthermore, we have
\begin{eqnarray*}
 A_m        &:=& b_{(m)}+\frac{1}{2} \Phi_m \bar\gamma-\frac{1}{2} \Phi_m^\dagger \gamma,\\
 b_{(m)}    &:=& \Upsilon_m b,\\
 \Upsilon_m &:=& \text{diag}(m,m-2,...,-m+2,-m),\\
 b          &:=& \frac{1}{2(R^2+y\bar y)}(\bar y~\dr y-y~\dr\bar y),\\
 \label{basechange-gamma-y}
 \gamma     &:=& \frac{\sqrt{2}R^2}{R^2+y\bar y}\dr y,\quad
 \bar \gamma :=  \frac{\sqrt{2}R^2}{R^2+y\bar y}\dr \bar y \,,\\
 \Phi_m &:=&
 \left(\begin{matrix}
	0&\phi_1&\cdots&0\\
	\vdots&0     &\ddots&\vdots\\
	\vdots&\vdots&\ddots&\phi_m\\
	0     &\cdots&\cdots&0
        \end{matrix}\right)\,,
\end{eqnarray*}
where also all $\phi_i$ are required to be real scalar fields on $\BR$. Here the 1-form $b$ is the gauge potential on the Dirac one-monopole line bundle over $\BC P^1$ and the $(1,0)$-form $\gamma$ as well as the $(0,1)$-form $\bar\gamma$ are the invariant basis of 1-forms on $\BC P^1$.

As one can easily see, the invariant 1-forms read
\begin{eqnarray*}
 \gamma 	&=& e^\gamma{}_y~\dr y,\\
 \bar\gamma 	&=& e^{\bar\gamma}{}_{\bar y}~\dr \bar y\,,
\end{eqnarray*}
where
\begin{equation*}
 \rho := e^\gamma{}_y = e^{\bar\gamma}{}_{\bar y}=\frac{\sqrt2 R^2}{R^2+y\bar y}\,.
\end{equation*}
The invariant metric $g$ on $\BC P^1\times \BC P^2$ is given by the non-vanishing components
\begin{subequations}
\begin{eqnarray}
 g_{a\bar b}=\de_{ab},~a,b\in\left\{1,2,\bar1,\bar2\right\},\\
 g_{y\by} = \rho^2,\text{ and hence } g^{\by y}=\rho^{-2}\,.
\end{eqnarray}
\end{subequations}
\paragraph{Maurer-Cartan equations and the field strength.}
We are dealing with invariant 1-forms on symmetric spaces and therefore these 1-forms fulfil the Maurer-Cartan equations, and can easily be calculated for the case of $\BC P^1$. The resulting equations for the invariant 1-forms are given by those of $\BC P^2$, written out in (\ref{maurcart-dB})-(\ref{maurcart-dbetatransp}), along with the ones corresponding to $\BC P^1$:
\begin{subequations}
\begin{eqnarray}
\label{maurcart-db}
 \dr b - \frac{1}{2R^2}~\bar\ga \wedge \ga &=& 0\,, \\
\label{maurcart-dgamma}
 \dr \ga - 2 b \wedge \ga & = & 0\,,\\
\label{maurcart-dbargamma}
 \dr \bg + 2 b \wedge \bg & = & 0\,.
\end{eqnarray}
\end{subequations}
If we insert the ansatz (\ref{ansatz cp1cp2 gauge potential}) into the definition $\CF=\dr \CA + \CA \wedge \CA$, we find the following field strength:
%
%\begin{footnotesize}
\begin{eqnarray}
 \label{ansatz cp1cp2 gauge field strength}
 \CF &=&\left(\rho^2 \left(\frac{1}{4} \left[\Phi_m^\top,\Phi_m\right] +\frac{1}{2R^2}\Y_m\right) \otimes \idm_3 \right)\dr\bar y\wedge\dr y %\nonumber \\
	+ \frac{\rho}{2}\dpar_t \Phi_m ~\dr t\wedge\dr\by
	- \frac{\rho}{2}\dpar_t \Phi_m^\top ~\dr t\wedge\dr y \nonumber\\
\nonumber \\
& &+	\frac{\rho}{2}\com{\Phi_m}{\Psi_m}
	   \otimes \left(\begin{matrix} 0_2&\dr\by \wedge \bar\beta \\ -\dr\by\wedge\beta^\top & 0 \end{matrix}\right)
	-\frac{\rho}{2}\com{\Phi_m^\top}{\Psi_m} 
	   \otimes \left(\begin{matrix} 0_2&\dr y \wedge \bar\beta \\ -\dr y\wedge\beta^\top & 0 \end{matrix}\right)
	\nonumber\\
& &+	\left(\Psi_m^2-\idm_{m+1}\right) \otimes
	\left(\begin{matrix}
   		 -\bb \wedge \beta^\top  & 0 \\ 0 & \beta^\dagger \wedge \be
   	\end{matrix}\right)
	+ \dpar_t(\Psi_m)\otimes
	\left(\begin{matrix}
		       0& \dr t \wedge \bb \\ -\dr t \wedge \be^\top & 0
		\end{matrix}\right)\,.
\end{eqnarray}
%\end{footnotesize}
% where
% \begin{eqnarray}
%  \left(\begin{matrix}
% 	\bar\beta^1\wedge\beta^1 & \bar\beta^1\wedge\beta^2\\
% 	\bar\beta^2\wedge \beta^1& \bar\beta^2\wedge \beta^2
%   \end{matrix}\right)
% &=& \bar\beta \wedge \beta^\top \,, \\
% (\bar\beta^1\wedge\beta^1+\bar\beta^2\wedge\beta^2)
% &=& \beta^\dagger\wedge \beta\,.
% \end{eqnarray}
%
\paragraph{Yang-Mills equations.}
The Yang-Mills equations on the space $\BR\times\BC P^1\times \BC P^2$ look slightly more complicated than before, since the dimension is higher than in the previous cases. We can see that, since we are dealing with a product of two projective spaces, we will get one more matrix equation for the additional $\BC P^1$. Namely, we have
\begin{eqnarray*}
 \CD_A \CF^{A\de} &=& 0, \qquad \de\in\left\{\by,~y\right\} \\
 \CD_A \CF^{A a} &=& 0, \qquad a \in \left\{1,~2,~\bar1~,\bar2\right\}
\end{eqnarray*}
which reads
\begin{equation}
\label{ansatz cp1cp2 yang-mills}
\begin{aligned}
 &\dpar_\tau\CF^{\tau\delta}&+&\nabla_\alpha \CF^{\alpha\delta}&+&\left[\CA_\alpha,\CF^{\alpha\delta}\right]&+
 &\nabla_c \CF^{c\delta}&+&\left[\CA_c,\CF^{c\delta}\right]& &=& &0& \,,\\
 &\dpar_\tau \CF^{\tau d}&+&\nabla_\alpha \CF^{\alpha d}&+&\left[\CA_\alpha,\CF^{\alpha d}\right]&+
 &\nabla_c \CF^{c d}&+&\left[\CA_c,\CF^{c d}\right]& &=& &0& \,.
\end{aligned}
\end{equation}
Here the repeated Greek indices are summed over the components belonging to $\BC P^1$, namely $\al\in\left\{y,~\by\right\}$, and the Latin letters are summed over the $\BC P^2$ components. The covariant derivatives of the field strength for the projective spaces are given in the canonical way for product spaces.

If we insert (\ref{ansatz cp1cp2 gauge potential}) and (\ref{ansatz cp1cp2 gauge field strength}) into (\ref{ansatz cp1cp2 yang-mills}), we arrive at the following matrix equations:
\begin{subequations}
\begin{eqnarray}
	\label{ansatz cp1cp2 matrix eom1}
 0 &=&
	\dpar_\tau^2 \Phi_m	
	-\frac{1}{4}\com{\Phi_m}{\com{\Phi_m^\top}{\Phi_m}}
	+\frac{1}{R^2}\Phi_m
	-\com{\Psi_m}{\com{\Phi_m}{\Psi_m}}\,,\\
	\label{ansatz cp1cp2 matrix eom2}
 0 &=& 	\dpar_\tau^2\Psi_m
	-\frac{1}{2}\com{\Phi_m^\top}{\com{\Phi_m}{\Psi_m}}
	+ 3~(\Psi_m-\Psi_m^3)\,.
\end{eqnarray}
\end{subequations}
Inserting $\Phi_m$ and $\Psi_m$ into these matrix equations (\ref{ansatz cp1cp2 matrix eom1}) and (\ref{ansatz cp1cp2 matrix eom2}), we find the following independent differential equations for our scalar fields $\psi_i$ and $\phi_i$ (denoting $\phi_{m+1}:=0=:\phi_0$):
\begin{subequations}
\label{ansatz cp1cp2 field eom}
\begin{eqnarray}
%\begin{aligned}
0 &=& 	\dpar_\tau^2\phi_i
	+\frac{1}{4}(\phi_{i-1}^2-2\phi_i^2+\phi_{i+1}^2)\phi_i
	+\frac{1}{R^2}\phi_i
	-\left(\psi_{i+1}^2-2\psi_{i+1}\psi_i+\psi_i^2\right)\phi_i \label{ansatz cp1cp2 field eom1}\,,\\
0 &=&   \dpar_\tau^2\psi_i-\frac{1}{2}\left(\phi_{i-1}^2(\psi_{i}-\psi_{i-1})+\phi_{i}^2(\psi_{i}-\psi_{i+1})\right) + 3(\psi_{i}-\psi_{i}^3)\,. 
%\end{aligned}
\label{ansatz cp1cp2 field eom2}
\end{eqnarray}
\end{subequations}
\begin{figure}[ht]
\subfigure[Solutions of (\ref{ansatz cp1cp2 field eom m=1}) with the following initial values:\newline $\phi_1(0)=1,~\dot\phi_1(0)=0,~\psi_1(0)=1,~\dot\psi_1(0)=0,~
\newline \psi_2(0)=0,~\dot\psi_2(0)=1.3$
]{
  \includegraphics[width=7.9cm]{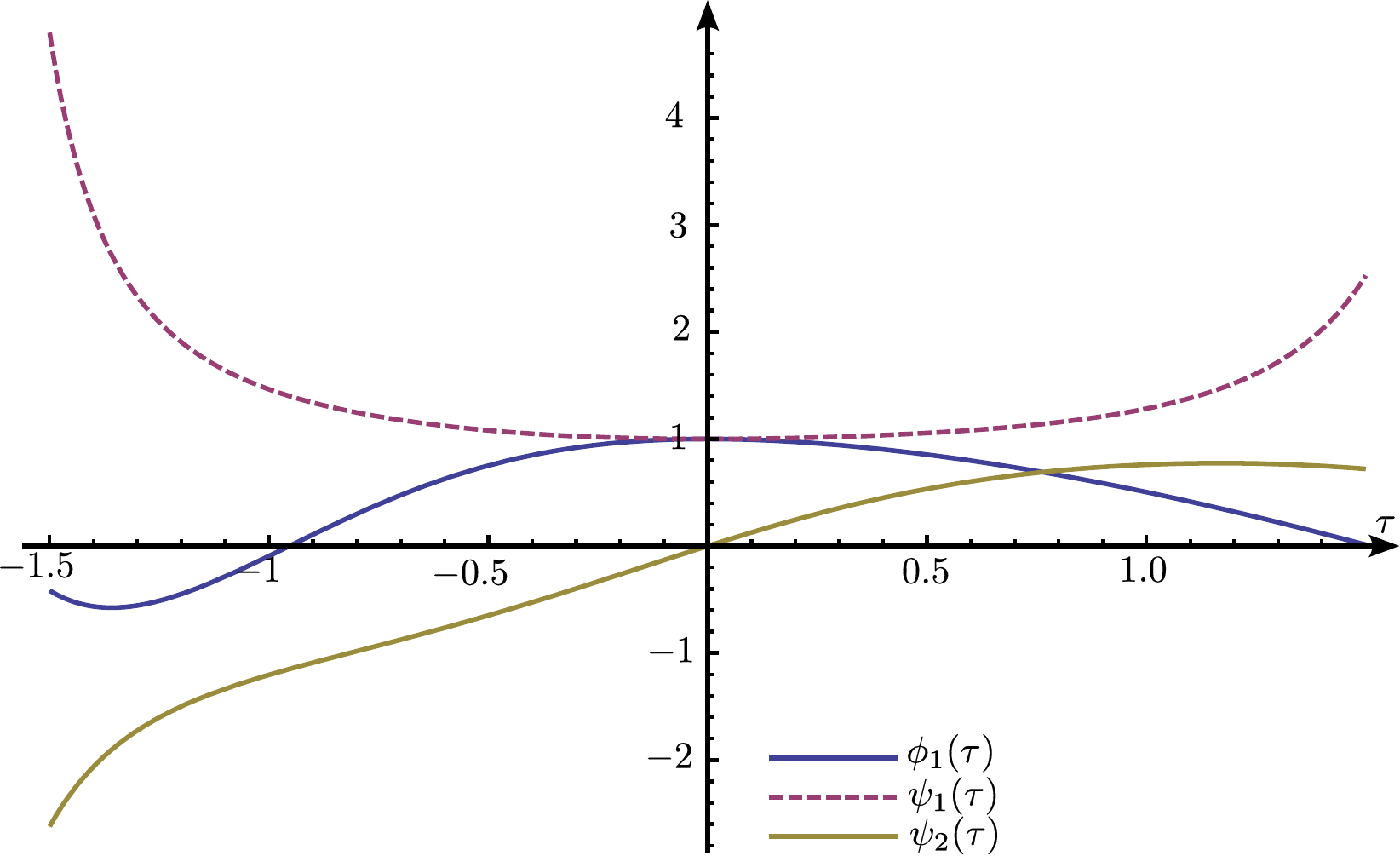}
}
\subfigure[Solutions of (\ref{ansatz cp1cp2 field eom m=1}) with the following initial values:\newline $\phi_1(0)=1,~\dot\phi_1(0)=0,~\psi_1(0)=0,~\dot\psi_1(0)=1,~
\newline \psi_2(0)=0,~\dot\psi_2(0)=0.5$
]{
  \includegraphics[width=7.9cm]{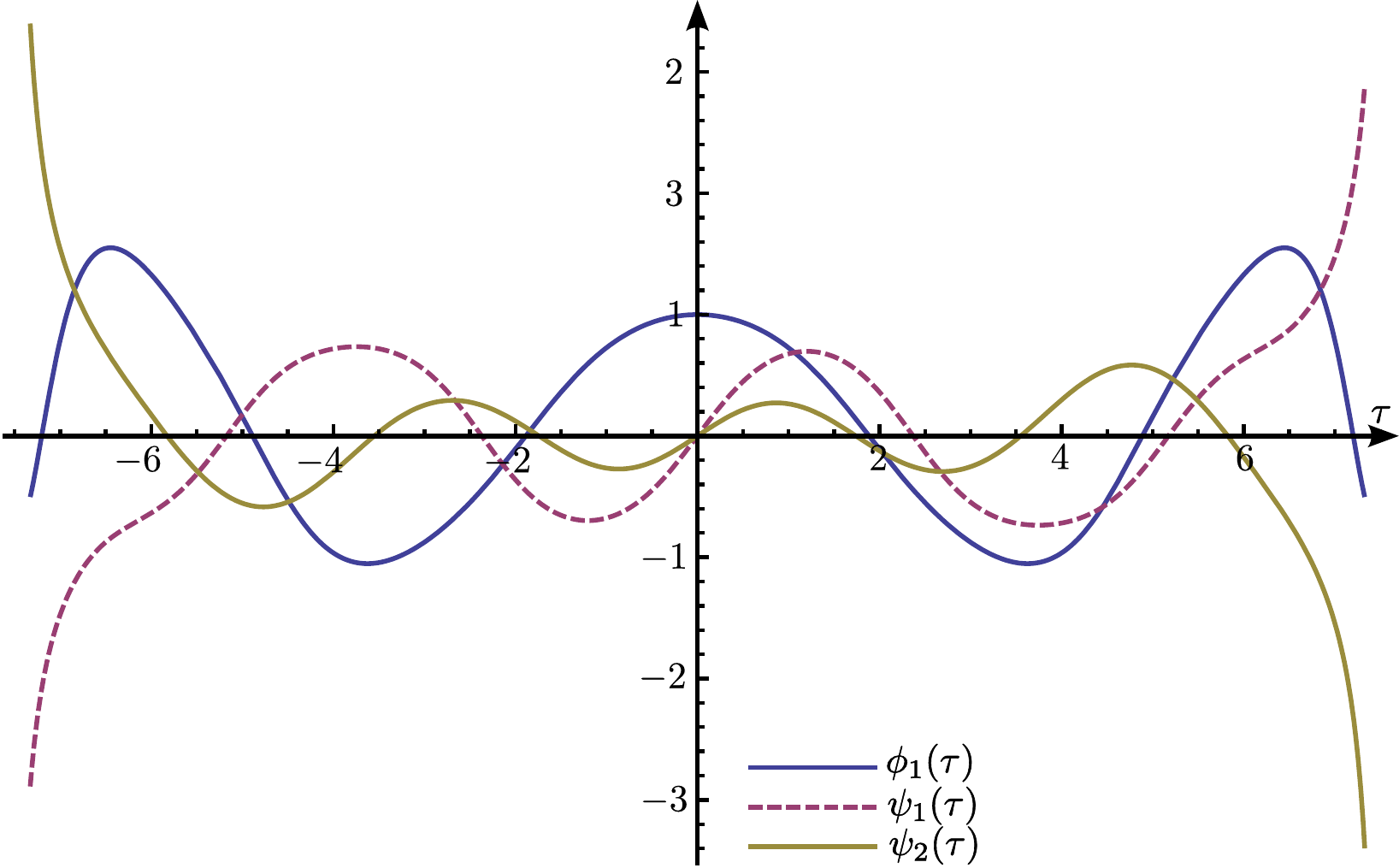}
}
  \caption{\small Two different sets of truly coupled numerical solutions $\phi_1(\tau),~\psi_1(\tau),~\psi_2(\tau)$ to the system of differential equations (\ref{ansatz cp1cp2 field eom m=1}). Particularly $\psi_1\neq\psi_2,~,\psi_i\neq\phi_1,~\psi_i\neq0,~\phi_1\neq0,~\forall~i\in\{1,2\} $.}
  \label{fig:RxCP1xCP2}
\end{figure}
Solutions to similar equations were found in \cite{preprint2} and one may in principle construct some solutions of (\ref{ansatz cp1cp2 field eom1})-(\ref{ansatz cp1cp2 field eom2}). For instance, for $m=1$ our system (\ref{ansatz cp1cp2 field eom}) reduces to three equations
\begin{subequations}
\label{ansatz cp1cp2 field eom m=1}
\begin{eqnarray}
0 &=& \dpar_\tau^2\phi_1 + \phi_1\left(\frac{1}{R^2}+\left(\psi_1-\psi_2\right)^2\right)-\frac12 \phi_1^3\\
\label{ansatz cp1cp2 field eom1 m=1}
0 &=& \dpar_\tau^2\psi_1 + 3\left(\psi_1-\psi_1^3\right) + \frac12 \phi_1^2 \left(\psi_2-\psi_1\right)\\
\label{ansatz cp1cp2 field eom2 m=1}
0 &=& \dpar_\tau^2\psi_2 + 3\left(\psi_2-\psi_2^3\right) + \frac12 \phi_1^2 \left(\psi_1-\psi_2\right)\,.
\label{ansatz cp1cp2 field eom3 m=1}
\end{eqnarray}
\end{subequations}

Here we can see that for instance by putting $\psi_1=\psi_2$ the system of differential equations decouples and one could write down the specific solutions for the three Higgs fields as we did in previous sections. But also as before it is not easy to write down the truly coupled solutions for this system. Nevertheless we were still able to provide some numerical solutions to (\ref{ansatz cp1cp2 field eom m=1}) which you can find in figure \ref{fig:RxCP1xCP2}.
\section*{ACKNOWLEDGMENTS}
\addcontentsline{toc}{section}{Acknowledgments}
I would like to thank Alexander D.\ Popov for substantial support during the development of this work. I would also like to thank Tatiana A.\ Ivanova, Kirsten Vogeler and Olaf Lechtenfeld for helpful discussions and remarks. This work was done within the framework of the project supported by the Deutsche Forschungsgemeinschaft under the grant 436 RUS 113/995. 
%
%\bibliography{Bibliography.bib}

\begin{thebibliography}{10}

\bibitem{CDFN}
{E.\ Corrigan, C.\ Devchand, D.\ B.\ Fairlie and J.\ Nuyts}, ``{First-Order
  Equations for Gauge Fields in Spaces of Dimension Greater than Four},''
\href{http://dx.doi.org/10.1016/0550-3213(83)90244-4}{{\em {Nucl.\ Phys.\ B}}
  {\bf {214}} ({1983})  {452}}.
%%CITATION = NUPHA,B214,452;%%.

\bibitem{BPS2}
{R.\ S.\ Ward}, ``{Complete Solvable Gauge-Field Equations in Dimensions
  Greater than Four},''
\href{http://dx.doi.org/10.1016/0550-3213(84)90542-X}{{\em {Nucl.\ Phys.\ B}}
  {\bf {236}} ({1984})  {381}}.
%%CITATION = NUPHA,B236,381;%%.

\bibitem{GSW}
{M.\ B.\ Green, J.\ H.\ Schwarz and E.\ Witten}, {\em {Superstring Theory}}.
\newblock {Cambridge University Press}, {1987}.

\bibitem{fairly-1984}
{D.\ B.\ Fairlie and J.\ Nuyts}, ``{Spherically Symmetric Solutions of Gauge
  Theories in Eight Dimensions},''
\href{http://dx.doi.org/10.1088/0305-4470/17/14/030}{{\em {J.\ Phys.\ A}} {\bf
  {17}} ({1984})  {2867}}.
%%CITATION = JPAGB,A17,2867;%%.

\bibitem{fubini-1985}
{S.\ Fubini and H.\ Nicolai}, ``{The Octonionic Instanton},''
\href{http://dx.doi.org/10.1016/0370-2693(85)91589-8}{{\em {Phys.\ Lett.\ B}}
  {\bf {155}} ({1985})  {369}}.
%%CITATION = PHLTA,B155,369;%%.

\bibitem{ivanova-1992}
{T.\ A.\ Ivanova and A.\ D.\ Popov}, ``{Self-Dual Yang-Mills Fields in d = 7,
  8, Octonions and Ward Equations},''
\href{http://dx.doi.org/10.1007/BF00402672}{{\em {Lett.\ Math.\ Phys.\ }} {\bf
  {24}} ({1992})  {85}}.
%%CITATION = LMPHD,24,85;%%.

\bibitem{ivanova-1993}
{T.\ A.\ Ivanova and A.\ D.\ Popov}, ``{(Anti)Self-Dual Gauge Fields in
  Dimension $d\geq 4$},''
\href{http://dx.doi.org/10.1007/BF01019334}{{\em {Theor.\ Math.\ Phys.\ }} {\bf
  {94}} ({1993})  {225}}.
%%CITATION = TMPHA,94,225;%%.

\bibitem{preprint1}
{A.\ D.\ Popov}, ``{Explicit Non-Abelian Monopoles and Instantons in SU(N) Pure
  Yang-Mills Theory},''
  \href{http://dx.doi.org/10.1103/PhysRevD.77.125026}{{\em {Phys.\ Rev.\ }}
  {\bf {D77}} ({2008})  {125026}},
\href{http://arxiv.org/abs/0803.3320}{{\tt arXiv:0803.3320 [hep-th]}}.
%%CITATION = 0803.3320;%%.

\bibitem{preprint2}
{A.\ D.\ Popov}, ``{Bounces/Dyons in the Plane Wave Matrix Model and SU(N)
  Yang-Mills Theory},'' \href{http://dx.doi.org/10.1142/S0217732309030163}{{\em
  {Mod.\ Phys.\ Lett.\ A}} {\bf {24}} ({2009})  {349}},
\href{http://arxiv.org/abs/0804.3845}{{\tt arXiv:0804.3845 [hep-th]}}.
%%CITATION = 0804.3845;%%.

\bibitem{preprint3}
{T.\ A.\ Ivanova and O.\ Lechtenfeld}, ``{Yang-Mills Instantons and Dyons on
  Group Manifolds},''
  \href{http://dx.doi.org/10.1016/j.physletb.2008.10.027}{{\em {Phys.\ Lett.\
  B}} {\bf {670}} ({2008})  {91}},
\href{http://arxiv.org/abs/0806.0394}{{\tt arXiv:0806.0394 [hep-th]}}.
%%CITATION = 0806.0394;%%.

\bibitem{yangmillsflow}
{T.\ A.\ Ivanova, O.\ Lechtenfeld, A.\ D.\ Popov and T.\ Rahn}, ``{Instantons
  and Yang-Mills Flows on Coset Spaces},''
\href{http://dx.doi.org/10.1007/s11005-009-0336-1}{{\em {Lett.\ Math.\ Phys.\ }} {\bf
  {89}} ({2009})  {231}},
\href{http://arxiv.org/abs/0904.0654}{{\tt arXiv:0904.0654 [hep-th]}}.
%%CITATION = 0904.0654;%%.

\bibitem{CSDR}
{D.\ Kapetanakis and G.\ Zoupanos}, ``{Coset Space Dimensional Reduction Of
  Gauge Theories},''
\href{http://dx.doi.org/10.1016/0370-1573(92)90101-5}{{\em {Phys.\ Rept.\ }} {\bf {219}} ({1992})  {1}}.
%%CITATION = PRPLC,219,1;%%.

\bibitem{math.DG/0112160}
{L.\ \'Alvarez-C\'onsul and O.\ Garcia-Prada}, ``{Dimensional Reduction and
  Quiver Bundles},'' {\em {J.\ Reine Angew.\ Math.\ }} {\bf {556}} ({2003})  1,
  \href{http://arxiv.org/abs/arXiv:math/0112160v2}{{\tt arXiv:math/0112160v2}}.

\bibitem{preprint6}
{O.\ Lechtenfeld, A.\ D.\ Popov and R.\ J.\ Szabo}, ``{Quiver Gauge Theory and
  Noncommutative Vortices},''
  \href{http://dx.doi.org/10.1143/PTPS.171.258}{{\em {Prog.\ Theor.\ Phys.\
  Suppl.\ }} {\bf {171}} ({2007})  {258}},
\href{http://arxiv.org/abs/0706.0979}{{\tt arXiv:0706.0979 [hep-th]}}.
%%CITATION = 0706.0979;%%.

\bibitem{preprint4}
{A.\ D.\ Popov and R.\ J.\ Szabo}, ``{Quiver Gauge Theory of Nonabelian
  Vortices and Noncommutative Instantons in Higher Dimensions},''
  \href{http://dx.doi.org/10.1063/1.2157005}{{\em {J.\ Math.\ Phys.}} {\bf
  {47}} ({2006})  {012306}},
\href{http://arxiv.org/abs/hep-th/0504025}{{\tt arXiv:hep-th/0504025}}.
%%CITATION = HEP-TH/0504025;%%.

\bibitem{rank2quiver}
{O.\ Lechtenfeld, A.\ D.\ Popov and R.\ J.\ Szabo}, ``{Rank Two Quiver Gauge
  Theory, Graded Connections and Noncommutative Vortices},''
  \href{http://dx.doi.org/10.1088/1126-6708/2006/09/054}{{\em {JHEP}} {\bf
  {09}} ({2006})  054},
\href{http://arxiv.org/abs/hep-th/0603232}{{\tt arXiv:hep-th/0603232}}.
%%CITATION = HEP-TH/0603232;%%.

\bibitem{Szabo:2009-1}
{B.\ P.\ Dolan and R.\ J.\ Szabo}, ``{Dimensional Reduction, Monopoles and
  Dynamical Symmetry Breaking},''
  \href{http://dx.doi.org/10.1088/1126-6708/2009/03/059}{{\em {JHEP}} {\bf
  {03}} ({2009})  {059}},
\href{http://arxiv.org/abs/0901.2491}{{\tt arXiv:0901.2491 [hep-th]}}.
%%CITATION = 0901.2491;%%.

\bibitem{Szabo:2009-2}
{B.\ P.\ Dolan and R.\ J.\ Szabo}, ``{Dimensional Reduction and Vacuum
  Structure of Quiver Gauge Theory},''
 \href{http://iopscience.iop.org/1126-6708/2009/08/038/}{{\em {JHEP}} {\bf
  {08}} ({2009})  {038}},
\href{http://arxiv.org/abs/0905.4899}{{\tt arXiv:0905.4899 [hep-th]}}.
%%CITATION = 0905.4899;%%.

\bibitem{preprint5}
{O.\ Lechtenfeld, A.\ D.\ Popov and R.\ J.\ Szabo}, ``{SU(3)-Equivariant Quiver
  Gauge Theories and Nonabelian Vortices},''
  \href{http://dx.doi.org/10.1088/1126-6708/2008/08/093}{{\em {JHEP}} {\bf 08}
  (2008)  093},
\href{http://arxiv.org/abs/0806.2791}{{\tt arXiv:0806.2791 [hep-th]}}.
%%CITATION = 0806.2791;%%.

\bibitem{Louis:2001}
{J.\ Louis and A.\ Micu}, ``{Heterotic String Theory with Background Fluxes},''
  \href{http://dx.doi.org/10.1016/S0550-3213(02)00040-8}{{\em Nucl. Phys.} {\bf
  B626} (2002)  26},
\href{http://arxiv.org/abs/hep-th/0110187}{{\tt arXiv:hep-th/0110187}}.
%%CITATION = HEP-TH/0110187;%%.

\bibitem{Cardoso:2002}
{G.\ L.\ Cardoso, G.\ Curio, G.\ Dall'Agata, D.\ L\"ust, P.\ Manousselis and
  G.\ Zoupanos}, ``{Non-K\"ahler String Backgrounds and Their Five Torsion
  Classes},'' \href{http://dx.doi.org/10.1016/S0550-3213(03)00049-X}{{\em Nucl.
  Phys.} {\bf B652} (2003)  5},
\href{http://arxiv.org/abs/hep-th/0211118}{{\tt arXiv:hep-th/0211118}}.
%%CITATION = HEP-TH/0211118;%%.

\bibitem{Cardoso:2003}
{G.\ L.\ Cardoso, G.\ Curio, G.\ Dall'Agata and D.\ L\"ust}, ``{BPS Action and
  Superpotential for Heterotic String Compactifications with Fluxes},''
  \href{http://dx.doi.org/10.1088/1126-6708/2003/10/004}{{\em JHEP} {\bf 10}
  (2003)  004},
\href{http://arxiv.org/abs/hep-th/0306088}{{\tt arXiv:hep-th/0306088}}.
%%CITATION = HEP-TH/0306088;%%.

\bibitem{Frey:2005}
{A.\ R.\ Frey and M.\ Lippert}, ``{AdS Strings with Torsion: Non-Complex
  Heterotic Compactifications},''
  \href{http://dx.doi.org/10.1103/PhysRevD.72.126001}{{\em Phys. Rev.} {\bf
  D72} (2005)  126001},
\href{http://arxiv.org/abs/hep-th/0507202}{{\tt arXiv:hep-th/0507202}}.
%%CITATION = HEP-TH/0507202;%%.

\bibitem{Chatzistavrakidis:2008}
{A.\ Chatzistavrakidis, P.\ Manousselis and G.\ Zoupanos}, ``{Reducing the
  Heterotic Supergravity on Nearly-Kahler Coset Spaces},''
  \href{http://dx.doi.org/10.1002/prop.200900012}{{\em Fortschr. Phys.} {\bf
  57} (2009)  527--534},
\href{http://arxiv.org/abs/0811.2182}{{\tt arXiv:0811.2182 [hep-th]}}.
%%CITATION = 0811.2182;%%.

\bibitem{Chatzistavrakidis:2009}
{A.\ Chatzistavrakidis and G.\ Zoupanos}, ``{Dimensional Reduction of the
  Heterotic String over Nearly-K\"ahler Manifolds},''
\href{http://dx.doi.org/10.1088/1126-6708/2009/09/077}{{\em JHEP} {\bf 09}
  (2009)  001},
\href{http://arxiv.org/abs/0905.2398}{{\tt arXiv:0905.2398 [hep-th]}}.
%%CITATION = 0905.2398;%%.

\end{thebibliography}

\providecommand{\href}[2]{#2}\begingroup\raggedright\endgroup

\addcontentsline{toc}{section}{References}

\end{document}